\documentclass[a4paper,11pt]{article}
\pdfoutput=1 

\usepackage{jheppub} 

\usepackage[T1]{fontenc} 
\usepackage{tikz} 
\usetikzlibrary{calc,through,backgrounds,patterns,decorations.pathmorphing,decorations.markings,%
		trees,positioning,arrows,chains,shapes.geometric,%
    decorations.pathreplacing,shapes}

\usepackage{amsmath}
\usepackage{multirow}
\usepackage{array}

\usepackage{enumerate}

\newcommand{\CF}{\mathcal{F}}
\def\CJ{\mathcal{J}}

\numberwithin{equation}{section}

\title{\boldmath   Holographic D3-probe-D5 Model of a Double Layer Dirac Semimetal}

\author[a ]{Gianluca Grignani,}
\author[b]{Namshik Kim,}
\author[a ]{Andrea Marini,} 
\author[b]{Gordon W. Semenoff}
   
\affiliation[a]{Dipartimento di Fisica e Geologia, Universit\`a di Perugia,
INFN Sezione di Perugia, 
Via A. Pascoli, 06123 Perugia, Italia}
\affiliation[b]{ Department of Physics and Astronomy, University of British Columbia,
6224 Agricultural Road, Vancouver, British Columbia, Canada V6T 1Z1}

\emailAdd{ grignani@pg.infn.it}
\emailAdd{namshik@phas.ubc.ca}
\emailAdd{marini@pg.infn.it}
\emailAdd{gordonws@phas.ubc.ca }

\abstract{
The possibility of inter-layer exciton condensation in 
a holographic D3-probe-D5 brane model  of a strongly coupled 
double monolayer Dirac semi-metal in a magnetic field is studied in detail.  
It is found that, when the charge densities on the layers are exactly balanced so that, at weak coupling, the Fermi surfaces
of electrons in one monolayer and holes in the other monolayer would be perfectly nested, 
inter-layer condensates can form for any separation of the layers. The case where
both monolayers are charge neutral is special.  
There, the inter-layer condensate occurs only for small separations and is 
replaced by an intra-layer exciton condensate at larger separations. 
The phase diagram for charge balanced monolayers for a range layer separations and chemical potentials is found. 
 We also show that, in semi-metals with multiple species of massless fermions,  the balance of charges required for 
 Fermi surface nesting can occur spontaneously by breaking some of the internal symmetry of the monolayers. This could
 have important consequences for experimental attempts to find inter-layer condensates. }

\begin{document} 
\maketitle
\flushbottom

  \section{Introduction and Summary}
\label{sec:intro}

The possibility of Coulomb drag-mediated exciton condensation in double monolayer graphene 
or other multi-layer heterostructures has recently received considerable attention \cite{bilayercondensate1}-\cite{genoa}. 
The term ``double monolayer graphene'' refers to two monolayers of graphene\footnote{
It should be distinguished from bilayer graphene where electrons are allowed to hop between the layers. }, each of which would be a Dirac semi-metal in isolation, 
and which are brought into 
close proximity but are still separated by an insulator so that direct transfer of electric charge carriers between the layers is negligible.   
The system then has two conserved charges, the electric charge in each layer.   
The Coulomb interaction between an electron in one layer and a hole in the other layer is attractive.    
A bound state of an electron and a hole that forms due to this attraction is called an exciton.  Excitons are bosons and, at low temperatures they
can condense to a form a charge-neutral superfluid.    
We will call this an inter-layer exciton condensate.  
Electrons and holes in the same monolayer can also form an exciton bound state, which we will call this an intra-layer exciton and its
Bose condensate an intra-layer condensate.

Inter-layer excitons  have   been observed in some cold atom analogs of double monolayers \cite{gaas1}-\cite{gaas3}
and as a 
transient phenomenon  in Gallium Arsenide/Aluminium-Gallium-Arsenide double quantum wells, 
albeit only at low temperatures and in the presence of  magnetic fields\cite{gaas3}-\cite{gaas5}. 
Their study is clearly of  interest for understanding fundamental issues 
with quantum coherence over mesoscopic distance scales and dynamical symmetry breaking.
 Recent interest in this possibility in graphene double layers
has been inspired by some theoretical modelling  which seemed to indicate 
that the exciton condensate could occur at room temperature \cite{bilayercondensate2}.  
A room temperature superfluid would have interesting applications 
in electronic devices where proposals include ultra-fast switches and 
dispersionless field-effect transistors \cite{device}-\cite{device5}.   This has motivated some recent experimental studies of double  
monolayers of graphene separated
by ultra-thin insulators, down to the nanometer scale \cite{man1}\cite{kim}. 
These experiments have revealed interesting features of
the phenomenon of Coulomb drag.  However, to this date, 
coherence between monolayers has yet to be observed in a stationary state
of a double monolayer.

One impediment to a truly quantitative analysis of  inter-layer coherence is the fact
that the Coulomb interaction at sub-nanoscale distances  is strong and perturbation theory must be re-summed in 
an ad hoc way to take screening
into account \cite{bilayercondensate3}\cite{screen}.  
In fact, inter-layer coherence will likely always require strong interactions.  The purpose of this paper is
to point out the existence of an inherently nonperturbative model of  very strongly coupled multi-monolayer systems.
This model is a defect quantum field theory which is the holographic AdS/CFT dual  of a  D3-probe-D5 brane system. 
 It is simple to analyze and exactly solvable in the limit where the quantum field theory
interactions are strong. External magnetic field and charge density can be incorporated into the solution and 
it exhibits a rich phase diagram where it has phases with inter-layer exciton condensates.

It might be expected that, with a sufficiently strong attractive electron-hole interaction, an inter-layer condensate would always form.   
One of the  lessons
of our work will be that this is not necessarily so.  In fact, it was already suggested in reference \cite{Evans:2013jma} that, when both
monolayers are charge neutral, and in a constant external magnetic field,
there can be an inter-layer or an intra-layer condensate but there were no phases where the two kinds of condensate  both occur at the same time.
What is more, the inter-layer condensate only appears for small separations of the monolayers, up to a critical separation.
As the spacing between the monolayers
is increased to the critical distance, there is a phase transition where an intra-layer condensate takes over.   
Intra-layer condensates in a strong magnetic field are already well known to occur in monolayer graphene in the integer quantum Hall regime
\cite{newhallplateaus}-\cite{newhallplateaus3}. They are thought to be a manifestation of  
``quantum Hall ferromagnetism''  \cite{quantumhallferromagnet0}-\cite{quantumhallferromagnet6}  or the ``magnetic catalysis of chiral symmetry
breaking'' \cite{cat0}-\cite{Filev:2012ch}   
which involve symmetry
breaking with an intra-layer exciton condensate.  It has been argued that the latter 
phenomenon, intra-layer exciton condensation, in a single monolayer 
is also reflected in symmetry breaking behaviour of the D3-probe-D5 brane system 
\cite{Kristjansen:2012ny}-\cite{Kristjansen:2013hma}.

 Another striking conclusion  that we will come to is that, even in the strong coupling limit, there is no inter-layer exciton condensate unless the
charge densities of the monolayers are fine-tuned in such a way that, at weak coupling, the electron Fermi surface on one monolayer
and the hole Fermi surface in the other monolayer are perfectly nested, that is, they have identical Fermi energies.   
In this particle-hole symmetric theory, this means that the charge
densities on the monolayers are of equal magnitude and opposite sign.  It is surprising that this need for charge balance
is even sharper in the strong coupling limit
than what is seen at weak coupling, where the infrared singularity from nesting does provide the instability 
needed for exciton condensation, but where, also,  
there is a narrow window near perfect nesting where condensation is still possible \cite{bilayercondensate1.6}. 
In our model, at strong coupling, there is inter-layer condensate only in the perfectly nested (or charge balanced) case. 
This need for such fine tuning  of charge densities could help to explain why such a condensate is hard to 
find in experiments where charged impurities would disturb the charge balance.  

\begin{figure}[!ht]
	\begin{center}
	\includegraphics[scale=1.2]{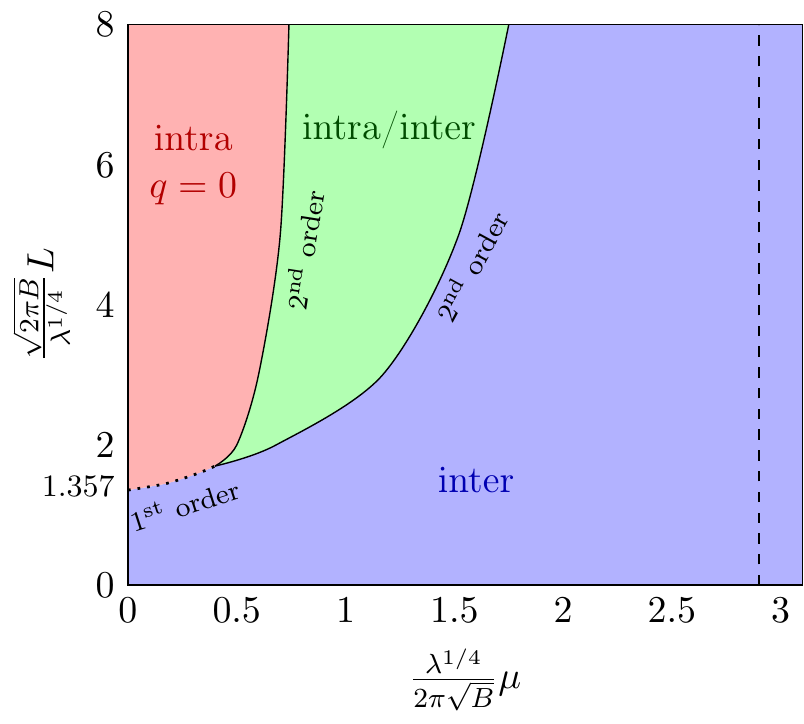}
		\caption{Phase diagram of the D3-probe-D5 brane system with balanced charge densities.  Layer separation  is plotted on the vertical axis
	and the chemical potential $\mu$  for electrons in one monolayer and holes 
	in the other monolayer is plotted on the horizontal axis.  The units employed set the length scale
	$\sqrt{ \frac{\sqrt{\lambda}}{ 2\pi B}} $ equal to one.   
	The blue region is a phase with an inter-layer condensate and 
	with no intra-layer condensate.  The green region is a phase with both an inter-layer and an intra-layer condensate.  The red
	region has only an intra-layer condensate.  In that region, the chemical potential is too small to induce a density of the 
	massive electrons ( $\mu$ is in the charge gap) and the charge densities on both of the monolayers vanishes.   The electrons and holes are 
	massive in that phase due to the intra-layer exciton condensate. The dotted line, separating a pure inter-layer from a pure intra-layer
	condensate, is a line of first order phase transitions.   The solid lines, on the other hand, indicate second order transitions.
	 		}
			\label{fig:phased0}	
\end{center}
\end{figure}

 When the charge densities of the monolayers are non-zero, and when they are balanced, 
 there can be an inter-layer condensate at any separation of the monolayers.  The phase diagram which
 we shall find for the D3-probe-D5 brane system in a magnetic field and with nonzero, balanced charge densities is
 depicted in figure \ref{fig:phased0}.   The blue region has an inter-layer condensate and no intra-layer condensate.  
 The green region has both inter-layer and intra-layer condensates.  The red region has only an intra-layer condensate.
 From the vertical axis in figure \ref{fig:phased0} we see that, in the charge neutral case. the inter-layer condensate exists
 only for separation less than a critical one.    
 
It has recently been suggested \cite{GKMS1} that there is another possible behaviour which can lead to 
inter-layer condensates when the charges of the  monolayers are not balanced.    This can occur when the material of the monolayers
contain more than one species of 
fermions.   For example, graphene has four species and emergent SU(4) symmetry \cite{Semenoff:1984dq}.   
In that case, the most symmetric state of a monolayer has the charge
of that monolayer shared equally by each of the four species of electrons.   Other less symmetric states are possible.  

Consider, for example, 
the double monolayer with one monolayer having  electron charge density $Q$ and the other monolayer having
hole density $\bar Q$ (or electron charge density $-\bar Q$), with    $Q>\bar Q>0$.   
On the hole-charged monolayer, some subset, which must be one, two or three
of the fermion species could take up all of the hole charge density, $\bar Q$. Then,  in the electron monolayer, the same number, one, two or three species
of electrons would take up electron charge density $\bar Q$ and the remainder of the species
will take electron charge density $Q-\bar Q$.   The (one, two or three) species with matched charge densities
will then spawn an inter-layer exciton condensate.  The remaining species on the hole monolayer is charge neutral.  A charge-neutral 
monolayer will have an intra-layer condensate.  The remaining
species in the electron monolayer, with charge density $Q-\bar Q$, would also have an intra-layer condensate and it would not have a charge gap
(all of the fermions are massive, but this species has a finite density and it  does not have a charge gap).
A simple signature of this state would be that one of the monolayers is charge gapped, whereas the other one is not.  
   The implication is that perfect fine-tuning of Fermi surfaces is not
absolutely necessary for inter-layer condensation.  We will show that, in a few examples, this type of spontaneous nesting can occur.
However, some important questions, 
such as how unbalanced the charge densities can 
be so that there is still a condensate are left for future work.

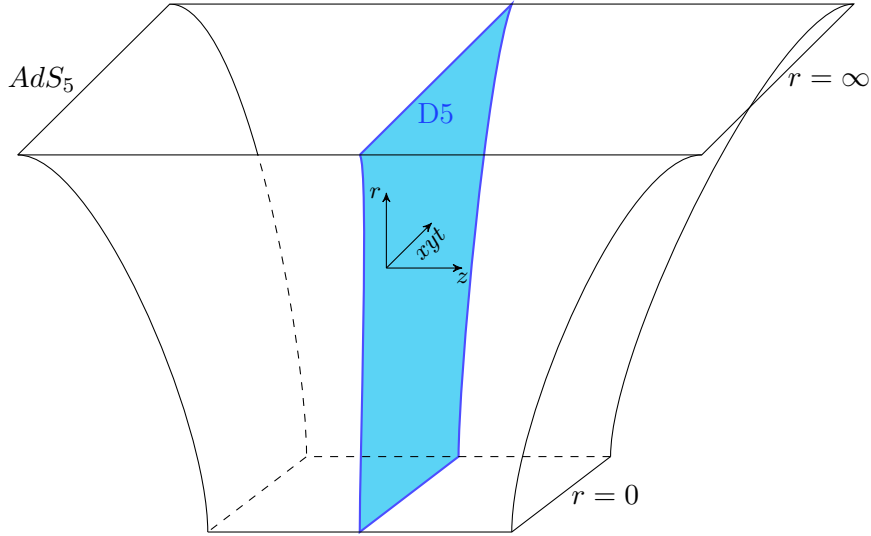
\begin{figure}[h!]

	\begin{center}
		\begin{tikzpicture}[>=stealth']
		\def\lunga{4.5};
		\def\larga{2};
		\def\lungb{2};
		\def\largb{1};
		\def\altez{5};
		\coordinate (A1) at (-\lunga,\altez);
		\coordinate (B1) at (\lunga,\altez);
		\coordinate (C1) at (2+\lunga,\altez+\larga);
		\coordinate (D1) at (2-\lunga,\altez+\larga);
		\coordinate (A2) at (-\lungb,0);
		\coordinate (B2) at (\lungb,0);
		\coordinate (C2) at (1.3+\lungb,\largb);
		\coordinate (D2) at (1.3-\lungb,\largb);
		\coordinate (D5a) at (0,\altez);
		\coordinate (D5b) at (2,\altez+\larga);
		\coordinate (D5c) at (1.3,\largb);
		\coordinate (D5d) at (0,0);
		\draw[blue,thick,fill=cyan!70,opacity=0.7] (D5a) --  (D5b) node[midway,below=5pt] {D5} .. controls +(70:-1) and +(90:1) .. (D5c) -- (D5d) .. controls +(90:1) and +(130:-0.2) .. (D5a);
		\draw (A1) .. controls +(0:1) and +(90:1.5) ..  (A2);
		\draw (B1) .. controls +(0:-1) and +(90:1.5) ..  (B2);
		\draw (C1) .. controls +(0:-1) and +(90:1.5) ..  (C2);
		\begin{scope}
			\clip (A1) rectangle (B2);
			\draw[dashed] (D1) .. controls +(0:1) and +(90:1.5) ..  (D2);
		\end{scope}
		\begin{scope}
			\clip (D1) rectangle (B1);
			\draw (D1) .. controls +(0:1) and +(90:1.5) ..  (D2);
		\end{scope}
		\draw (A1) --  (B1)  -- (C1) node[midway,right] {$r=\infty$} -- (D1) -- (A1) node[midway,left=3pt] {$AdS_5$};
		\draw (A2) --  (B2)  -- (C2) node[midway,right] {$r=0$};
		\draw [dashed] (C2) -- (D2) -- (A2);
		\draw[->] (.35,3.5) -- +(0,1) node[left=-2pt] {\footnotesize $r$};
		\draw[->] (.35,3.5) -- +(.6,.6) node[near end,below=-2pt,sloped] {\footnotesize $xyt$};
		\draw[->] (.35,3.5) -- +(1,0) node[,below=-2pt] {\footnotesize $z$};
		    \end{tikzpicture}
   \end{center}
\caption{\small A D5 brane is embedded in  $AdS_5\times S^5$ where the metric of $AdS_5$ is 
$ds^2=\sqrt{\lambda}\alpha'[\frac{dr^2}{r^2}+r^2(dx^2+dy^2+dz^2-dt^2)]$ and the D5 brane world-volume
is an $AdS_4$ subspace which fills  $r,x,y,t$ and sits at a point in $z$.  The
$AdS_5$ boundary  is located at $r=\infty$ and the Poincar\`e horizon at $r=0$. The D5 brane also wraps a maximal, 
contractible $S^2$ subspace of 
 $S^5$ of the $AdS_5\times S^5$ background.  The internal bosonic symmetries of the configuration are $SO(3)$ of the wrapped $S^2$ and a 
further $SO(3)$ symmetry of the position of the maximal $S^2$ in $S^5$.  }
\label{figure1}
\end{figure}

 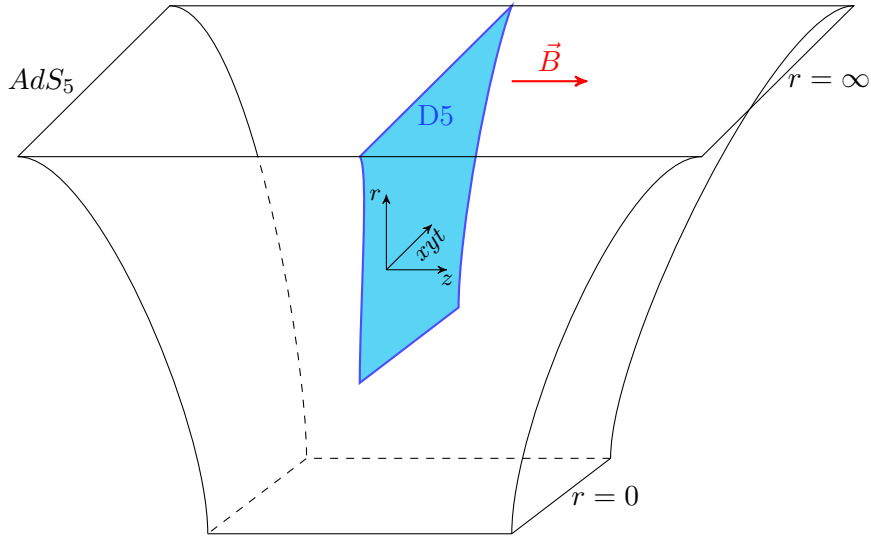
\begin{figure}[h!]
	\begin{center}
		\begin{tikzpicture}[>=stealth']
		\def\lunga{4.5};
		\def\larga{2};
		\def\lungb{2};
		\def\largb{1};
		\def\altez{5};
		\coordinate (A1) at (-\lunga,\altez);
		\coordinate (B1) at (\lunga,\altez);
		\coordinate (C1) at (2+\lunga,\altez+\larga);
		\coordinate (D1) at (2-\lunga,\altez+\larga);
		\coordinate (A2) at (-\lungb,0);
		\coordinate (B2) at (\lungb,0);
		\coordinate (C2) at (1.3+\lungb,\largb);
		\coordinate (D2) at (1.3-\lungb,\largb);
		\coordinate (D5a) at (0,\altez);
		\coordinate (D5b) at (2,\altez+\larga);
		\coordinate (D5c) at (1.3,2+\largb);
		\coordinate (D5d) at (0,2);
		\draw[blue,thick,fill=cyan!70,opacity=0.7] (D5a) --  (D5b) node[midway,below=5pt] {D5}  .. controls +(70:-.9) and +(90:1) .. (D5c) -- (D5d) .. controls +(90:1) and +(130:-0.2) .. (D5a);
		\draw (A1) .. controls +(0:1) and +(90:1.5) ..  (A2);
		\draw (B1) .. controls +(0:-1) and +(90:1.5) ..  (B2);
		\draw (C1) .. controls +(0:-1) and +(90:1.5) ..  (C2);
		\begin{scope}
			\clip (A1) rectangle (B2);
			\draw[dashed] (D1) .. controls +(0:1) and +(90:1.5) ..  (D2);
		\end{scope}
		\begin{scope}
			\clip (D1) rectangle (B1);
			\draw (D1) .. controls +(0:1) and +(90:1.5) ..  (D2);
		\end{scope}
		\draw (A1) --  (B1)  -- (C1) node[midway,right] {$r=\infty$} -- (D1) -- (A1) node[midway,left=3pt] {$AdS_5$};
		\draw (A2) --  (B2)  -- (C2) node[midway,right] {$r=0$};
		\draw [dashed] (C2) -- (D2) -- (A2);
		\draw[->] (.35,3.5) -- +(0,1) node[left=-2pt] {\footnotesize $r$};
		\draw[->] (.35,3.5) -- +(.6,.6) node[near end,below=-2pt,sloped] {\footnotesize $xyt$};
		\draw[->] (.35,3.5) -- +(.8,0) node[,below=-2pt] {\footnotesize $z$};
		\draw[->,thick,red] (2,6) -- +(1,0) node[midway,above] { $\vec{B}$};
    \end{tikzpicture}
   \end{center}
\caption{\small When the D5 brane is exposed to a magnetic field, it pinches off before it reaches the Poincar\`e horizon.  It does so for
any value of the magnetic field, with the radius at which it pinches off proportional to $\sqrt{\frac{2\pi B}{\sqrt{\lambda} }}$.  In this configuration, the embedding 
of the $S^2\subset S^5$ depends on the $AdS_5$ radius.   It is still 
the maximal one which can be embedded in $S^5$ at the boundary, but it shrinks and collapses to a point at the radius where the D5 brane
pinches off.}
\label{figure2}
\end{figure}

We will model a double monolayer system with a relativistic defect quantum field theory consisting of two parallel, infinite, planar 
2+1-dimensional defects embedded in 3+1-dimensional Minkowski space.  The defects are separated by a length, $L$.
 Some U(1) charged degrees of freedom inhabit  the defects and 
  play the role of the two dimensional relativistic electron gases.  We can consider states with charge densities on the monolayers.  
  As well, we can expose them to a constant external magnetic field.   We could also turn on a temperature and study them in a thermal
  state, however, we will not do so in this paper.  
  
  The theory that we use has an AdS/CFT dual, the D3-probe-D5 brane
system where the D5 and anti-D5 branes are probes embedded in the $AdS_5\times S^5$ background of the type IIB superstring theory.  
The $AdS_5\times S^5$ is sourced by $N$ D3 branes and it is tractable in the large $N$ limit where we simultaneously scale
the string theory coupling constant $g_s$ to zero so that $\lambda\equiv g_sN/4\pi=g_{YM}^2N$ is held constant. Here, $g_{\rm YM}$ is the coupling
constant of the gauge fields in the defect quantum field theory.   
The D5 and anti-D5 branes are semi-classical when the quantum field theory on the double monolayer is strongly coupled,
that is, where $\lambda$ is large.  It is solved by embedding a  
D5 brane and an anti-D5 brane in the  $AdS _5\times S^5$ background. 
The boundary conditions of the embedding are such that, as they approach the boundary of $AdS_5$, 
the world volumes approach the two parallel 2+1-dimensional monolayers.   The dynamical equations which we shall use are
identical for the brane and the anti-brane.   The reason why we use a brane-anti-brane pair is  that they can partially annihilate.  This
annihilation will be the string theory dual of the formation of an inter-layer exciton condensate.

The phase diagram of a single D5 brane or a stack of coincident D5 branes is well known \cite{Evans:2010hi}, with an important modification in
the integer quantum Hall regime \cite{Kristjansen:2012ny},\cite{Kristjansen:2013hma}. 
In the absence of a magnetic field or charge density, 
a single  charge neutral D5 brane takes up a supersymmetric and conformally invariant configuration.     
The D5 brane world-volume is itself $AdS_4$ and it stretches from the boundary of $AdS_5$ 
to the Poincar\`e horizon, as depicted in figure \ref{figure1}.   It also wraps an $S^2\subset S^5$.  
This is a maximally symmetric solution of the theory. It has a well-established quantum field theory dual
whose Lagrangian is known explicitly \cite{Karch:2000gx}-\cite{Erdmenger:2002ex}. 
The latter is a conformally symmetric phase of a defect super-conformal quantum field theory \footnote{Of course a supersymmetric conformal field
theory is not a realistic model of a semimetal.   Here,  we will use this model with a strong magnetic field.   It was observed in references \cite{Kristjansen:2012ny}, 
  \cite{Kristjansen:2013hma}  that the supersymmetry
and conformal symmetry are both broken by an external magnetic field, and that the low energy states of the weakly coupled system were
states with partial fillings of the fermion zero modes which occur in the magnetic field (the charge neutral point Landau level). 
The dynamical problem to be solved is that of  deciding which partial fillings of zero modes have the lowest energy.  It is  a
direct analog of the same problem in graphene or other Dirac semimetals.   It is in this regime that D3-D5 system exhibits quantum Hall 
ferromagnetism and other interesting
phenomena which can argued to be a strong coupling extrapolation of 
universal features of a semimetal in a similar environment.  
It is for this reason that we will concentrate on the system with a magnetic field, with the assumption that the very low energy states  of 
the theory are the most important for the physics of exciton formation, and that this situation persists to strong coupling.  There have been a number 
of works which have used D branes to model double monolayers \cite{Davis:2011am}-\cite{Filev:2014bna}.  }.

 Now, let us introduce  a magnetic field on the D5 brane world volume.  This is dual to the 2+1-dimensional field theory in a background
constant magnetic field.  
As soon as an external magnetic field is introduced, the single D5 brane changes its geometry drastically \cite{Filev:2009xp}.  
The brane pinches off and truncates at a finite $AdS_5$-radius, 
before it reaches the Poincar\`e horizon.  This is called a ``Minkowski embedding'' and is depicted in figure \ref{figure2}.   
This configuration has a charge gap. Charged  degrees of freedom 
are open strings which stretch from the D5 brane to the Poincar\`e horizon. When, 
the D5 brane does not reach the Poincar\`e horizon, these
strings have a minimum length and therefore a 
mass gap.  This is the gravity dual of the mass generation that accompanies exciton condensation in a single monolayer.

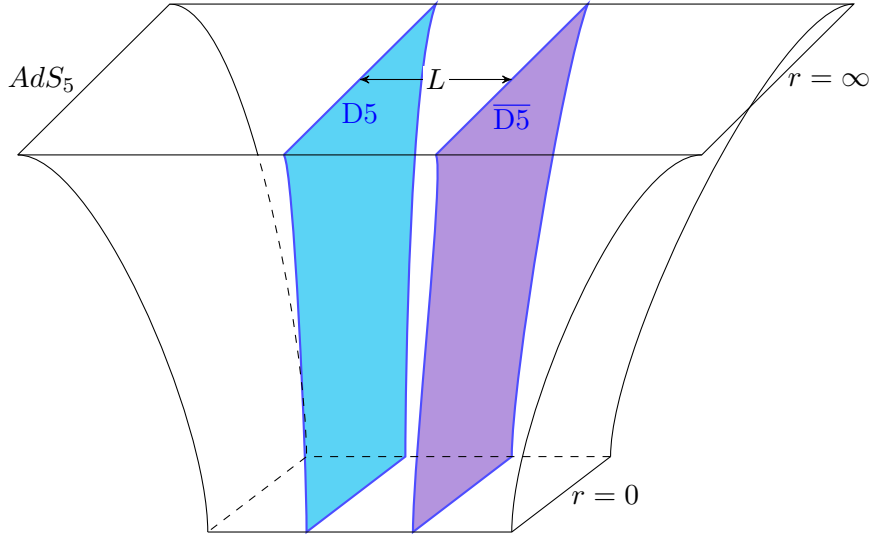
\begin{figure}[h!]
	\begin{center}
		\begin{tikzpicture}[>=stealth']
		\def\lunga{4.5};
		\def\larga{2};
		\def\lungb{2};
		\def\largb{1};
		\def\altez{5};
		\coordinate (A1) at (-\lunga,\altez);
		\coordinate (B1) at (\lunga,\altez);
		\coordinate (C1) at (2+\lunga,\altez+\larga);
		\coordinate (D1) at (2-\lunga,\altez+\larga);
		\coordinate (A2) at (-\lungb,0);
		\coordinate (B2) at (\lungb,0);
		\coordinate (C2) at (1.3+\lungb,\largb);
		\coordinate (D2) at (1.3-\lungb,\largb);
		\coordinate (D5a) at (-1,\altez);
		\coordinate (D5b) at (1,\altez+\larga);
		\coordinate (D5c) at (.6,\largb);
		\coordinate (D5d) at (-.7,0);
		\coordinate (D5aa) at (1,\altez);
		\coordinate (D5bb) at (3,\altez+\larga);
		\coordinate (D5cc) at (2,\largb);
		\coordinate (D5dd) at (.7,0);
		\draw[blue,thick,fill=cyan!70,opacity=0.7] (D5a) --  (D5b) node[midway,below=5pt,opacity=1] { D5}  .. controls +(70:-1) and +(90:1) .. (D5c) -- (D5d) .. controls +(90:1) and +(130:-0.2) .. (D5a);
		\draw[blue,thick,fill=blue!70!red!60,opacity=0.7] (D5aa) --  (D5bb) node[midway,below=5pt,opacity=1] {$\overline{\mathrm{D5}}$}  .. controls +(70:-1) and +(90:1) .. (D5cc) -- (D5dd) .. controls +(90:1) and +(130:-0.2) .. (D5aa);
		\draw (A1) .. controls +(0:1) and +(90:1.5) ..  (A2);
		\draw (B1) .. controls +(0:-1) and +(90:1.5) ..  (B2);
		\draw (C1) .. controls +(0:-1) and +(90:1.5) ..  (C2);
		\begin{scope}
			\clip (A1) rectangle (B2);
			\draw[dashed] (D1) .. controls +(0:1) and +(90:1.5) ..  (D2);
		\end{scope}
		\begin{scope}
			\clip (D1) rectangle (B1);
			\draw (D1) .. controls +(0:1) and +(90:1.5) ..  (D2);
		\end{scope}
		\draw (A1) --  (B1)  -- (C1) node[midway,right] {$r=\infty$} -- (D1) -- (A1) node[midway,left=3pt] {$AdS_5$};
		\draw (A2) --  (B2)  -- (C2) node[midway,right] { $r=0$};
		\draw [dashed] (C2) -- (D2) -- (A2);
		\draw [<->] (0,6) -- + (2,0) node[midway,fill=white,inner sep=1pt] {$L$};
		    \end{tikzpicture}
   \end{center}
\caption{\small A D5 brane and an anti-D5 brane are are suspended in $AdS_5$ as shown.  They are held a distance $L$ apart
at the $AdS_5$ boundary. }
\label{figure4}
\end{figure}

Let us now consider the double monolayer system.  We will begin with the case where both of the monolayers are charge neutral and there 
is no magnetic field.
We will model the strong coupled system by a pair which consists of a probe D5 brane and a probe anti-D5 brane suspended in the $AdS_5$ background
as depicted in figure \ref{figure4}.  Like a particle-hole pair, the D5 brane and the anti-D5 brane have a 
tendency to annihilate.  However, we can impose boundary conditions which prevent their annihilation.  
We require that,  as the D5 brane approaches the boundary
of the $AdS_5$ space, it is parallel to the anti-D5 brane and it is separated from the anti-D5 brane by a distance $L$.  
 Then, as each  brane  hangs down into the bulk of $AdS_5$, 
they  can still lower their energy by partially annihilating as depicted in the joined configuration in figure \ref{figure4}.    
This joining of the brane and anti-brane is the AdS/CFT dual of inter-layer exciton condensation.
The in this case, when they are both charge neutral, the branes will join for any value of the separation  $L$.  In this  strongly coupled defect quantum field
theory, with vanishing magnetic field and vanishing charge density on both monolayers,  the inter-layer exciton condensate exists for any value of the inter-layer distance.

\begin{figure}[h!]
	\begin{center}
		\begin{tikzpicture}[>=stealth']
		\def\lunga{4.5};
		\def\larga{2};
		\def\lungb{2};
		\def\largb{1};
		\def\altez{5};
		\coordinate (A1) at (-\lunga,\altez);
		\coordinate (B1) at (\lunga,\altez);
		\coordinate (C1) at (2+\lunga,\altez+\larga);
		\coordinate (D1) at (2-\lunga,\altez+\larga);
		\coordinate (A2) at (-\lungb,0);
		\coordinate (B2) at (\lungb,0);
		\coordinate (C2) at (1.3+\lungb,\largb);
		\coordinate (D2) at (1.3-\lungb,\largb);
		\coordinate (D5a) at (-1,\altez);
		\coordinate (D5b) at (1,\altez+\larga);
		\coordinate (D5c) at (1.4,2.2+\largb);
		\coordinate (D5d) at (.4,2.3);
		\coordinate (D5aa) at (1,\altez);
		\coordinate (D5bb) at (3,\altez+\larga);
		\draw[blue,thick,top color=cyan!80, bottom color=blue!60!cyan,opacity=0.7] (D5d) .. controls +(40:-.8) and  +(80:-1) ..  (D5a)   -- (D5b) node[midway,below=5pt,opacity=1] { D5} .. controls +(90:-1) and +(40:-1.7) .. (D5c); 
		\draw[blue,thick,top color=blue!70!red!60, bottom color=blue!60!cyan,opacity=0.7] (D5d) .. controls +(40:.3)  and +(90:-1.2) .. (D5aa)  -- (D5bb) node[midway,below=5pt,opacity=1] { $\overline{\mathrm{D5}}$} .. controls +(90:-1.5) and +(40:.8) .. (D5c);
		\draw (A1) .. controls +(0:1) and +(90:1.5) ..  (A2);
		\draw (B1) .. controls +(0:-1) and +(90:1.5) ..  (B2);
		\draw (C1) .. controls +(0:-1) and +(90:1.5) ..  (C2);
		\begin{scope}
			\clip (A1) rectangle (B2);
			\draw[dashed] (D1) .. controls +(0:1) and +(90:1.5) ..  (D2);
		\end{scope}
		\begin{scope}
			\clip (D1) rectangle (B1);
			\draw (D1) .. controls +(0:1) and +(90:1.5) ..  (D2);
		\end{scope}
		\draw (A1) --  (B1)  -- (C1) node[midway,right] {$r=\infty$} -- (D1) -- (A1) node[midway,left=3pt] {$ AdS_5$};
		\draw (A2) --  (B2)  -- (C2) node[midway,right] {$r=0$};
		\draw [dashed] (C2) -- (D2) -- (A2);
		\draw [<->] (0,6) -- + (2,0) node[midway,fill=white,inner sep=0pt] {$L$};
		    \end{tikzpicture}
   \end{center}
\caption{\small When the D5 brane and an anti-D5 brane are suspended as shown, their natural tendency is to join together.  This is the 
configuration with the lowest energy when $L$ is fixed.   It is also the configuration which describes the quantum field theory with an inter-layer exciton condensate.}
\label{figure3}
\end{figure}
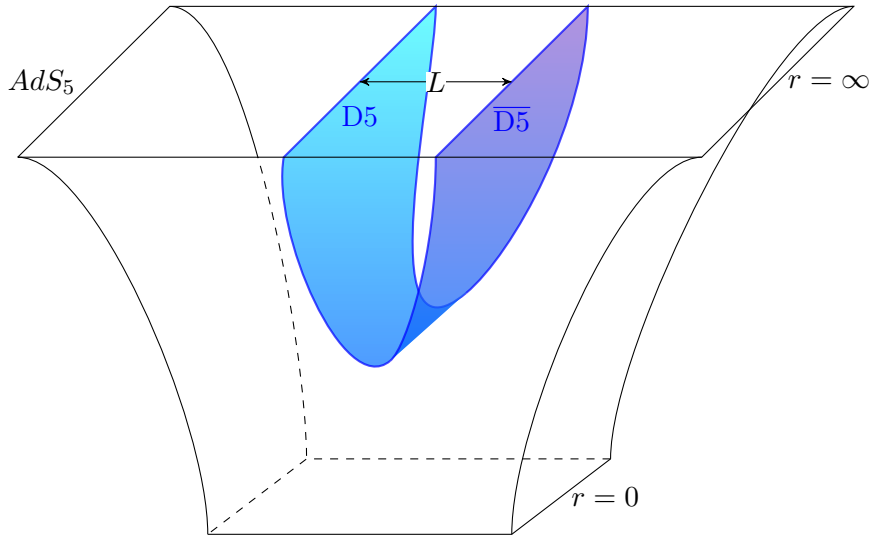

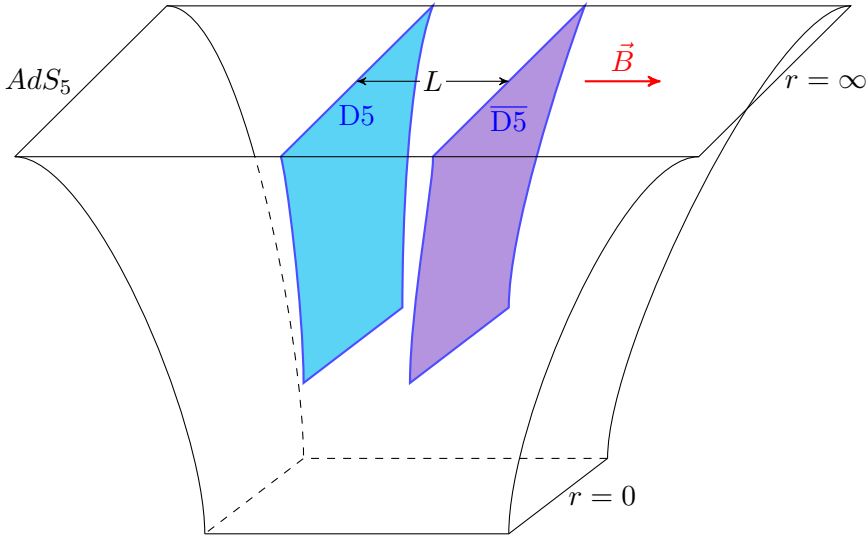
\begin{figure}[h!]
	\begin{center}
		\begin{tikzpicture}[>=stealth']
		\def\lunga{4.5};
		\def\larga{2};
		\def\lungb{2};
		\def\largb{1};
		\def\altez{5};
		\coordinate (A1) at (-\lunga,\altez);
		\coordinate (B1) at (\lunga,\altez);
		\coordinate (C1) at (2+\lunga,\altez+\larga);
		\coordinate (D1) at (2-\lunga,\altez+\larga);
		\coordinate (A2) at (-\lungb,0);
		\coordinate (B2) at (\lungb,0);
		\coordinate (C2) at (1.3+\lungb,\largb);
		\coordinate (D2) at (1.3-\lungb,\largb);
		\coordinate (D5a) at (-1,\altez);
		\coordinate (D5b) at (1,\altez+\larga);
		\coordinate (D5c) at (.6,2+\largb);
		\coordinate (D5d) at (-.7,2);
		\coordinate (D5aa) at (1,\altez);
		\coordinate (D5bb) at (3,\altez+\larga);
		\coordinate (D5cc) at (2,2+\largb);
		\coordinate (D5dd) at (.7,2);
		\draw[blue,thick,fill=cyan!70,opacity=0.7] (D5a) --  (D5b) node[midway,below=5pt,opacity=1] { D5}  .. controls +(70:-1) and +(90:1) .. (D5c) -- (D5d) .. controls +(90:1) and +(110:-0.2) .. (D5a);
		\draw[blue,thick,fill=blue!70!red!60,opacity=0.7] (D5aa) --  (D5bb) node[midway,below=5pt,opacity=1] {$\overline{\mathrm{D5}}$}  .. controls +(70:-1) and +(90:1) .. (D5cc) -- (D5dd) .. controls +(90:1) and +(100:-0.2) .. (D5aa);
		\draw (A1) .. controls +(0:1) and +(90:1.5) ..  (A2);
		\draw (B1) .. controls +(0:-1) and +(90:1.5) ..  (B2);
		\draw (C1) .. controls +(0:-1) and +(90:1.5) ..  (C2);
		\begin{scope}
			\clip (A1) rectangle (B2);
			\draw[dashed] (D1) .. controls +(0:1) and +(90:1.5) ..  (D2);
		\end{scope}
		\begin{scope}
			\clip (D1) rectangle (B1);
			\draw (D1) .. controls +(0:1) and +(90:1.5) ..  (D2);
		\end{scope}
		\draw (A1) --  (B1)  -- (C1) node[midway,right] {$r=\infty$} -- (D1) -- (A1) node[midway,left=3pt] {$AdS_5$};
		\draw (A2) --  (B2)  -- (C2) node[midway,right] { $r=0$};
		\draw [dashed] (C2) -- (D2) -- (A2);
		\draw [<->] (0,6) -- + (2,0) node[midway,fill=white,inner sep=1pt] {$L$};
		\draw[->,thick,red] (3,6) -- +(1,0) node[midway,above] {$\vec{B}$};
	
    \end{tikzpicture}
   \end{center}
\caption{\small When the D5 brane and an anti-D5 brane are exposed to a magnetic field, and if the field is strong enough, 
they can pinch off and end before they join.  
This tendency to pinch off competes with their tendency to join and in a strong enough field they will take up the phase that is shown where
they pinch off before they can join.}
\label{figure4a}
\end{figure}

If we now turn on a  magnetic field $B$ so that the dimensionless parameter  $BL^2$ is small, the branes   join as they did in the absence
of the field.  However, in a stronger field, as $BL^2$ is increased, there is a competition between the branes joining and, alternatively, 
each of the branes pinching off and truncating, as they would do if there were isolated.   The pinched off branes are depicted in figure \ref{figure4a}. 
This configuration has intra-layer exciton condensates on each monolayer but no inter-layer condensate.  
We thus see that, in a magnetic field,  the charge neutral double monolayer
always has a charge gap due to exciton condensation.   However, it has an inter-layer condensate only when the branes are close enough.

Now, we can also introduce a charge density on both the D5 brane and the anti-D5 brane.   We shall find a profound difference between the cases
where the overall density, the sum of the density on the two branes is zero and where it is nonzero.   In the first case, when it is zero, joined configurations
of branes exist for all separations.  Within those configurations, there are regions where the exciton condensate is inter-layer only and 
a region where it is a mixture of intra-layer and inter-layer.  These are seen in the phase diagram in figure  \ref{fig:phased0}.  The blue region has only an
inter-layer exciton condensate.  The green region has a mixed inter-layer and intra-layer condensate.   In the red region, the chemical potential is of
too small a magnitude to induce a charge density (it is in the charge gap) 
and the phase is identical to the neutral one, with an intra-layer condensate and no inter-layer condensate.  

In the case where the D5 and anti-D5 brane are not overall neutral, they cannot join.   There is never an inter-layer condensate.  They can have intra-layer
condensates if their separation is small enough.     However, there is another possibility, which occurs if we have stacks of multiple D5 branes.  In that case, 
there is the possibility that the D5 branes in a stack do not share the electric charge equally.   Instead some of them take on electric charges that matches
the charge of the anti-D5 branes, so that some of them can join, and the others absorb the remainder of the unbalanced charge and do not join.   At weak 
coupling this would correspond to a spontaneous nesting of the Fermi surfaces of some species of fermions in the monolayers, with the other species
taking up the difference of the charges.   At weak coupling, as well as in our strong coupling limit, the question is whether the spontaneously nested
system is energetically favored over one with a uniform distribution of charge.    We shall find that, for the few values of the charge where we have been able
to compare the energies, this is indeed the case.  

In the remainder of the paper, we will describe the quantitative analysis which leads to the above description of the behaviour of 
the D3-probe-D5 brane system. 
In section 2 we will discuss the mathematical problem of finding the geometry of probe D5 branes embedded in $AdS_5\times S^5$ in the 
configurations which give us the gravity dual of the double monolayer.   In section 3 we will discuss the behaviour of the double monolayer where
each layer is charge neutral and they are in 
a magnetic field.  In section 4 we will discuss the double monolayer in a magnetic field and with balanced charge densities.
In section 5 we will explore the behaviour of double monolayers with un-matched charge densities.  Section 6 contains some
discussion of the conclusions. 

\section{Geometry of branes with magnetic field and density}
\label{sec:gensetup}

We will consider a pair of probe branes, a D5 brane and an anti-D5 brane suspended in $AdS_5\times S^5$. 
They are both constrained to reach the boundary of $AdS_5$ with their world volume geometries approaching $AdS_4\times S^2$ where the $AdS_4$ is a subspace of $AdS_5$
 with one coordinate direction suppressed and $S^2$ is a maximal two-sphere embedded in $S^5$. What is more, 
  when they reach the boundary, we impose the boundary condition that they are separated from each other by a distance $L$. 
 
 We shall use coordinates where the metric of the $AdS_5\times S^5$ background is
\begin{equation}
ds^2=\frac{dr^2}{r^2}+r^2\left(-dt^2+dx^2+dy^2+dz^2\right)+d\psi^2+\sin^2\psi d^2\Omega_2+\cos^2\psi d^2\tilde{\Omega}_2
\end{equation}
where $d^2\Omega_2=d\theta^2+\sin^2\theta d\phi^2$ and $d^2\tilde \Omega_2=d\tilde\theta^2+\sin^2\tilde \theta d\tilde \phi^2$ 
are the metrics of two 2-spheres, $S^2$ and $\tilde S^2$. 
The world volume geometry of the D5 brane is for the most part determined by symmetry.   
We require Lorentz and translation invariance in 2+1-dimensions.   
This is achieved by both the D5  and the anti-D5 brane wrapping the subspace of $AdS_5$ with coordinates $t,x,y$.
We will also assume that all solutions have an SO(3) symmetry.  This is achieved when both the D5 and anti-D5 
brane world volumes wrap the 2-sphere $S^2$
with coordinates  $\theta,\phi$.  Symmetry requires that none of the remaining variables depend on $t,x,y,\theta,\phi$.  For the remaining internal coordinate
of the D5 or anti-D5 brane, it is convenient to use the projection of the $AdS_5$ radius, $r$ onto the brane world-volume.  
The D5 and anit- D5 branes will sit at points in the remainder of the $AdS_5\times S^5$ directions,  $z,\psi,\tilde\theta,\tilde\phi$.   The points  $z(r)$
and $ \psi(r)$ generally depend on $r$ and these functions become the dynamical variables of the embedding (along with world volume gauge fields
which we will introduce shortly). The wrapped $S^2$
has an $SO(3)$ symmetry.  What is more, 
the point $\psi=\frac{\pi}{2}$ where the wrapped sphere is maximal has an additional $SO(3)$ symmetry\footnote{At that point 
where $S^2$ is maximal, $\sin\psi=\sin\frac{\pi}{2}=1$ and $\cos\psi=0$, that is, the volume of $\tilde S^2$ vanishes.
The easiest way to see  that this embedding has an $SO(3)$ symmetry
is to parameterize the $S^5$ by $(x_1,...,x_6)$ with $x_1^2+\ldots x_6^2=1$.  $S^2$ is the space $x_1^2+x_2^2+x_3^2=\sin^2\psi$ 
and $\tilde S^2$ is $x_4^2+x_5^2+x_6^2=\cos^2\psi$. The point $\cos\psi=0$ with $x_4=x_5=x_6=0$ requires no choice of position on $\tilde S^2$ and it thus
has $SO(3)$ symmetry.   On the other hand, if $\cos\psi\neq 0$ and therefore 
some of the coordinates $(x_4,x_5,x_6)$ are nonzero, the symmetry is reduced to an $SO(2)$ rotation about the direction
chosen by the vector $(x_4,x_5,x_6)$. }.
The geometry of the D5 brane and the anti-D5 brane are both given by the ansatz 
\begin{equation}\label{worldvolumemetric}
ds^2=\frac{dr^2}{r^2}\left(1+(r^2 z')^2+(r\psi')^2\right)+r^2\left(-dt^2+dx^2+dy^2\right)+\sin^2\psi d^2\Omega_2
\end{equation}
The introduction of a charge density and external magnetic field will require D5 world-volume gauge fields. 
In the $a_r=0$ gauge, the field strength 2-form $F$ is given by 
\begin{equation}\label{worldvolumegaugefields}
\frac{2\pi}{\sqrt{\lambda}}F=a'_0(r)dr\wedge dt+b  dx\wedge dy
\end{equation}
In this expression, $b$ is a constant which will give a constant magnetic field in the holographic dual and $a_0(r)$ will
result in the world volume electric field which is needed in order to have a nonzero $U(1)$ charge density in the quantum
field theory.  The magnetic field $B$ and temporal gauge field $A_0$ are defined in terms of them as
\begin{equation}\label{physicalgaugefields}
b=\frac{2\pi}{\sqrt{\lambda}} B~,~~~~~a_0=\frac{2\pi}{\sqrt{\lambda}}A_0
\end{equation}
In this Section, we will use the field strength (\ref{worldvolumegaugefields}) for both the D5 brane and the anti-D5 brane.  

The asymptotic behavior  at $r\to\infty$ for the embedding functions in (\ref{worldvolumemetric})
and the gauge field (\ref{worldvolumegaugefields}) 
are such that
the sphere $S^2$ becomes maximal,
\begin{equation}\label{bcs}
\psi(r)\to \frac{\pi}{2}+\frac{c_1}{r}+\frac{c_2}{r^2}+\dots
\end{equation}
and the D5 brane and anti-D5 brane are separated by a distance $L$, 
\begin{align}\label{bcs1}
z(r)\to \frac{L}{2}-\frac{f}{r^5}+\dots
\end{align}
for the D5 brane and 
\begin{align}\label{bcs2}
z(r)\to -\frac{L}{2}+\frac{f}{r^5}+\dots
\end{align}
for the anti-D5 brane.  The asymptotic behaviour of the gauge field is 
\begin{align}
a_0(r)=\mu-\frac{q}{r}+\dots
\label{bcs3}
\end{align}
with $\mu$ and $q$ related to the chemical potential and the charge density, respectively.
There are two constants which specify the asymptotic behavior in each of the above equations. 
In all cases, we are free to choose one of the two constants as a boundary condition, for example 
we could choose $c_1,q,f$.  Then, the other
constants,  $c_2, \mu, L$, are fixed by requiring that the solution is non-singular.  

In this 
Paper, we will only consider solutions where the boundary condition is $c_1=0$.   This is the boundary condition that
is needed for the Dirac fermions in the double monolayer quantum field theory to be massless at the fundamental level.  Of course
they will not remain massless when there is an exciton condensate.  In the case where they are massless, we say that there is ``chiral symmetry'',
or that $c_1=0$ is a chiral symmetric
boundary condition.  Then, when we solve the equation of motion for $\psi(r)$, there are two possibilities.  
The first possibility is that
$c_2=0$ and $\psi=\frac{\pi}{2}$, a constant for all values of $r$.  This is the phase with good chiral symmetry.
Secondly, $c_2\neq 0$ and $\psi$ is a non-constant function of $r$.  This describes the phase with spontaneously broken
chiral symmetry.  The constant $c_2$ is proportional to the strength of the intra-layer chiral exciton condensate 
the D5 brane or the anti-D5 brane.  The constant $f$ instead is proportional to the strength of the inter-layer condensate.

To be more general, we could replace the single D5 brane by a stack of $N_5$ coincident D5 branes and the single anti-D5 brane
by another stack of $\bar N_5$ coincident anti-D5 branes.   Then, the main complication is that the world volume theories of the D5 and anti-D5
branes become non-abelian in the sense that the embedding coordinates become matrices and the world-volume gauge fields also have 
non-abelian gauge symmetry.  The Born-Infeld action must also be generalized to be, as well as an integral over coordinates, a trace over the 
matrix indices.   For now, we will assume that the non-abelian structure plays no significant role.  Then, all of the matrix degrees of freedom
are proportional to unit matrices and the trace in the non-abelian Born-Infeld action simply produces a factor of the number of branes, $N_5$ or $\bar N_5$
(see equation (\ref{S}) below).
We will also take $N_5=\bar N_5$ and leave the interesting possibility that $N_5\neq \bar N_5$ for future work.  (This generalization could, for example, describe 
the interesting situation where a double monolayer consists of a layer of
 graphene and a layer of topological insulator.)   We also have not searched for interesting non-Abelian solutions of the world volume theories which would
 provide other competing phases of the double monolayer system.   Some such phases are already known to exist.  For example, it was shown in references \cite{Kristjansen:2012ny} and 
 \cite{Kristjansen:2013hma} that, when the Landau level filling fraction, which is proportional to $Q/B$, is greater than approximately 0.5, there is a competing
 non-Abelian solution which resembles a D7 brane and which plays in important role in matching  there integer quantum Hall states which are
 expected to appear at integer filling fractions.   In the present work, we will avoid this region by assuming that the filling fraction is sub-critical. 
 Some other aspects of the non-Abelian structure will be important to us in section 5.

The Born-Infeld action for either the stack of D5 branes or the stack of anti-D5 branes is given by 
\begin{equation}\label{S}
S=\mathcal{N}_5\int dr\sin^2\psi \sqrt{r^4+b^2}\sqrt{1+(r\psi')^2+(r^2z')^2-(a'_0)^2}
\end{equation}
where 
\[
\mathcal{N}_5=\frac{\sqrt{\lambda} N N_5}{2\pi^3}V_{2+1}
\]
with $V_{2+1}$ the volume of the 2+1-dimensional space-time, $N$ the number of D3 branes, 
$N_5$ the number of D5 branes.  The Wess-Zumino terms that occur in the D brane action will not
play a role in the D5 brane problem. 

The variational problem of extremizing the Born-Infeld action (\ref{S}) involves two
cyclic variables, $a_0(r)$ and $z(r)$.  Being cyclic, their canonical momenta must be constants,
\begin{align}
&\label{Q}Q=-\frac{\delta S}{\delta A_0'}\equiv \frac{2\pi \mathcal{N}_5}{\sqrt{\lambda}} q~,~~~~q=\frac{\sin^2\psi\sqrt{r^4+b^2}a_0'}{\sqrt{1+(r\psi')^2+(r^2z')^2-(a'_0)^2}}\\
&\label{F} \Pi_z	=\frac{\delta S}{\delta z'}\equiv \mathcal{N}_5 f~,~~~~~~f=\frac{\sin^2\psi\sqrt{r^4+b^2}r^4z'}{\sqrt{1+(r\psi')^2+(r^2z')^2-(a'_0)^2}}~,
\end{align}

Solving \eqref{Q} and \eqref{F} for $a_0'(r)$ and $z'(r)$ in terms of $q$ and $f$ we get
\begin{equation}\label{a0p}
a'_0=\frac{q r^2 \sqrt{ 1+r^2 \psi '^2}}{ \sqrt{
   r^4 \left(b^2+r^4\right) \sin ^4\psi+q^2 r^4-f^2}}
\end{equation}
\begin{equation}\label{zp}
z'=\frac{f \sqrt{ 1+r^2 \psi '^2}}{r^2 \sqrt{
   r^4 \left(b^2+r^4\right) \sin ^4\psi+q^2 r^4-f^2}}
\end{equation}

The Euler-Lagrange equation can be derived by varying the action \eqref{S}.  We eliminate $a_0'(r)$ and $z'(r)$ from that equation
using equations \eqref{a0p} and \eqref{zp}.  Then  the equation of motion for $\psi$ reads
\begin{equation}\label{eompsi}
\frac{r \psi ''+ \psi '}{1+r^2 \psi '^2}-\frac{\psi ' \left(f^2+q^2 r^4+r^4 \left(b^2+3 r^4\right) \sin ^4\psi\right)-2 r^3
   \left(b^2+r^4\right) \sin ^3\psi \cos \psi }{f^2-q^2 r^4-r^4 \left(b^2+r^4\right) \sin ^4\psi}=0
\end{equation}
This equation must be solved with the boundary conditions in equation (\ref{bcs})-(\ref{bcs3})   (remembering that we can choose
only one of the integration constants, the other being fixed by regularity of the solution) 
in order to find the function $\psi(r)$.  Once we know that function, we can integrate
equations \eqref{a0p} and \eqref{zp} to find $a_0(r)$ and $z(r)$. 
 
Clearly, $\psi=\frac{\pi }{2}$, a constant, for all values of $r$,  is always a solution of equation (\ref{eompsi}), even when the magnetic field
and charge density are nonzero.   However, for some range 
of the parameters, it will not be the most stable solution.

\subsection{Length, Chemical Potential and Routhians}

The solutions of the equations of motion are implicitly functions of the integration constants.
We can consider a variation of the integration constants in such a way that the functions $\psi(r),a_0(r),z(r)$ 
remain solutions as the constants vary.   Then, the on-shell action varies in a specific way.  
Consider the action (\ref{S}) evaluated on solutions of 
the equations of motion.  We call the on-shell action the free energy ${\mathcal F}_1=S[\psi,z,a_0]/\mathcal{N}_5$.  
If we take  a variation of the parameters
in the solution, here, specifically $\mu$ and $L$, while keeping $c_1=0$, and assuming that
the equations of motion are obeyed, we obtain
\begin{align} 
\delta {\mathcal F}_1&= \int_0^\infty dr  \left( \delta \psi \frac{\partial{\mathcal L}}{\partial\psi'}+ 
\delta a_0 \frac{\partial{\mathcal L}}{\partial a_0'}+
\delta z \frac{\partial{\mathcal L}}{\partial z'}\right)'
\nonumber \\
&=- q\delta \mu +f\delta L  
\end{align}
The first term, with $\delta\psi$ vanishes because $\delta \psi\sim \delta c_2/r^2$. 
We see that ${\mathcal F}_1$ is  a function of the chemical potential $\mu$ 
and the distance $L$ and the conjugate variables, the charge density and the force
needed to hold the D5 brane and anti-D5 brane apart are gotten by taking 
partial derivatives,
\begin{align}
\left. q=-\frac{\partial {\mathcal F}_1}{\partial\mu} \right|_L~~,~~
\left. f=\frac{\partial{\mathcal F}_1}{\partial L}\right|_\mu
\end{align}
  When the dynamical system relaxes to its ground state, with the parameters
$\mu$ and $L$ held constant, it relaxes to a minimum of ${\mathcal F}_1$.   

There are other possibilities for free energies.  For example, the quantity which is minimum when
the charge
density, rather than the chemical potential, is fixed, is obtained from ${\mathcal F}_1[L,\mu]$ by a Legendre transform,
 \begin{align}\label{F2}
 \mathcal{F}_2[L,q]={\mathcal F}_1[L,\mu ]+q\mu
 \end{align}
 If we formally consider ${\mathcal F}_2$ off-shell as an action from which, for fixed $q$ and $f$, we can derive
 equations of motion for $\psi(r)$ and $z(r)$, 
 \begin{equation}\label{rou}
\mathcal{F}_2=\dfrac{S}{\mathcal{N}_5}+\int q {a'_0}dr=\int \,dr\sqrt{\sin^4\psi(r^4+b^2)+q^2}\sqrt{1+(r\psi')^2+(r^2z')^2}
\end{equation}
where  we have used
\[
a_0'=\frac{q \sqrt{1+r^4 z'^2+r^2 \psi '^2}}{\sqrt{\left(b^2+r^4\right) \sin ^4\psi+q^2}}
\]
obtained by solving equation  \eqref{Q}  for $a_0'$.
The equation of motion for $\psi(r)$, equation (\ref{eompsi}), can be derived from (\ref{rou}) by varying $\psi(r)$. 
Moreover, we still have  
\begin{align}\label{equationforz}
f=\frac{\sin^2\psi\sqrt{r^4+b^2}r^4z'}{\sqrt{1+(r\psi')^2+(r^2z')^2-(a'_0)^2}}=\frac{\sqrt{\left(b^2+r^4\right) \sin ^4\psi+q^2} \,r^4 z'}{\sqrt{1+r^4 z'^2+r^2 \psi '^2}}
\end{align}
which was originally derived from  (\ref{S}) by varying $z$ and then finding a first integral of the 
resulting equation of motion.  It can also be  derived from \eqref{rou}. 

Once the function $\psi(r)$ is known, we can solve equation (\ref{equationforz}) for $z'(r)$ and then integrate to
compute the separation of the D5 and anti-D5 branes, 
\begin{equation}\label{L}
L=2\int_{r_0}^\infty\, dr\, z'(r)=2f\int_{r_0}^{\infty}\, dr\,\frac{ \sqrt{ 1+r^2 \psi '^2}}{r^2 \sqrt{
   r^4 \left(b^2+r^4\right) \sin ^4\psi+q^2 r^4-f^2}}
\end{equation}
where $\psi(r)$ is a solution of \eqref{eompsi} and $r_0$ is the turning point, that is the place where the denominator in the integrand vanishes.  This turning point depends on the value of $\psi(r_0)$.
When $\psi$ is the constant solution $\psi=\pi/2$, 
\begin{equation}\label{r0}
r_0=\frac{\sqrt[4]{\sqrt{\left(b^2+q^2\right)^2+4 f^2}-b^2-q^2}}{\sqrt[4]{2}}
\end{equation}
and the integral in \eqref{L} can be done analytically. It reads
\begin{equation}\label{Lanalytic}
L=2 f\, \int_{r_0}^{\infty}\, dr\,\frac{1}{r^2 \sqrt{
   r^4 \left(b^2+r^4\right)+q^2 r^4-f^2}}=\frac{f\,\sqrt{\pi } \Gamma \left(\frac{5}{4}\right) \,
   _2F_1\left(\frac{1}{2},\frac{5}{4};\frac{7}{4};-\frac{f^2}{{r_0}^8}\right)}{2 {r_0}^5\Gamma\left(\frac{7}{4}\right) \sqrt{b^2+q^2}}
\end{equation}
For $b=q=0$, $f=r_0^4$, we get
\[
L=\frac{2 \sqrt{\pi } \Gamma \left(\frac{5}{8}\right)}{r_0\Gamma \left(\frac{1}{8}\right)}
\]
in agreement with  the result quoted in reference \cite{Evans:2013jma} .

Analogously, the chemical potential is related to the integral of the gauge field strength on the brane in the $(r,0)$ directions, \eqref{a0p},
\begin{equation}\label{mu}
\mu=\int_{r_0}^\infty a'_0(r)\,dr= q
\int_{r_0}^{\infty}\, dr\,\frac{r^2 \sqrt{ 1+r^2 \psi '^2}}{\sqrt{
   r^4 \left(b^2+r^4\right) \sin ^4\psi+q^2 r^4-f^2}}
   \end{equation}

When $\psi$ is the constant solution $\psi=\pi/2$ the integral in \eqref{mu} can again be done analytically and reads 
\begin{equation}\label{mupi2}
\mu=q
\int_{r_0}^{\infty}\, dr\, \frac{ r^2}{\sqrt{r^4 \left(b^2+r^4\right)+q^2 r^4-f^2}}
=\frac{q\,\sqrt{\pi } \Gamma \left(\frac{5}{4}\right) \,
   _2F_1\left(\frac{1}{4},\frac{1}{2};\frac{3}{4};-\frac{f^2}{{r_0}^8}\right)}{{r_0} \Gamma
   \left(\frac{3}{4}\right)}
\end{equation}

Through equations \eqref{L} and \eqref{mu}, $L$ and $\mu$ are viewed as functions of $f$ and $q$, this equations can in principle be inverted to have $f$ and $q$ as functions of $L$ and $\mu$.

We can now use \eqref{a0p} and \eqref{zp} to eliminate $a_0'$ and $z'$ from the action \eqref{S}
to get the expression of the free energy $\mathcal{F}_1$
\begin{equation}\label{SLmu}
{\mathcal F}_1[L,\mu]=\int_{r_0}^\infty \,dr\,  \left(b^2+r^4\right) \sin ^4\psi  \frac{r^2 \sqrt{ 1+r^2 \psi '^2}}{\sqrt{
   r^4 \left(b^2+r^4\right) \sin ^4\psi+q^2 r^4-f^2}}
\end{equation}
this has to be thought of as a function of $L$ and $\mu$, where $f$ and $q$ are considered as functions of $L$ and $\mu$, given implicitly by \eqref{L} and \eqref{mu}.
Note that we do   not  do a Legendre transform here since we need the variational functional which is a function of $L$ and $\mu$ the D5 brane separation and chemical potential that are the physically relevant parameters. 

Using \eqref{zp} to eliminate  $z'$ in the Routhian  \eqref{rou}, we now get a function of $L$ and $q$
\begin{equation}\label{R1}
\mathcal{F}_2[L,q]=\int_{r_0}^\infty  \,dr\, \left(\left(b^2+r^4\right) \sin ^4\psi+q^2\right) \frac{r^2 \sqrt{ 1+r^2 \psi '^2}}{\sqrt{r^4 \left(b^2+r^4\right) \sin ^4\psi+q^2 r^4-f^2}}
   \end{equation}
 The Routhian \eqref{R1} is a function of $L$  through the fact that it is a function of $f$ and $f$ is a function of $L$ given implicitly by \eqref{L}.
Of course had we performed the Legendre transform of the Routhian also with respect to $L$, the result would be
\begin{equation}\label{R2}
\mathcal{F}_3[f,q]=\mathcal{F}_2[L,q]-\int f z'dr
=\int_{r_0}^\infty  \,dr\,  \frac{\sqrt{ 1+r^2 \psi '^2}}{r^2}\sqrt{r^4 \left(b^2+r^4\right) \sin ^4\psi+q^2 r^4-f^2}
   \end{equation}
which is the variational functional appropriate for variations which hold both $q$ and $f$ fixed.

Note that, for convenience, from now on we shall scale the magnetic field $b$ to 1 
in all the equations and formulas we wrote: This can be easily done implementing the following rescalings 
\begin{equation}\label{rescbq}
	r\to \sqrt{b} r\,,\quad f\to b^2 f\, , \quad q \to b \, q\, , \quad L \to \sqrt{b}L\, , \quad \mu \to \frac{\mu}{\sqrt{b}}\, , \quad \CF_i\to b^{3/2}\CF_i \, .
\end{equation}

\section{Double monolayers with a magnetic field}

In reference \cite{Evans:2013jma} the case of a double monolayer where both of the monolayers are charge neutral was considered 
with an external magnetic field.     In this section, we will re-examine their results within our framework and using our notation.
The equation of motion for $\psi(r)$ in this case is
\begin{equation}\label{qzero}
\frac{r \psi ''+ \psi '}{1+r^2 \psi '^2}-\frac{\psi ' \left(f^2+r^4 \left(1+3 r^4\right) \sin ^4\psi\right)-2 r^3
   \left(1+r^4\right) \sin ^3\psi \cos \psi }{f^2-r^4 \left(1+r^4\right) \sin ^4\psi}=0
\end{equation}

There are in principle four type of solutions for which $c_1=0$ in \eqref{bcs}~\cite{Evans:2013jma}:
\begin{enumerate}
\item An unconnected, constant solution that reaches the Poincar\'e horizon.  An embedding of the D5 brane which reaches
the Poincar\'e horizon is called a ``black hole (BH) embedding''.  Being a constant solution, this  corresponds to a  
 state of the double monolayer where both the intra-layer and inter-layer condensates vanish. 
 \item A connected constant $\psi=\frac{\pi}{2}$ solution. Since this is a connected solution, $z(r)$ has a non trivial profile in $r$ 
 and its  boundary behaviour is given by equation \eqref{bcs2} with $f$ non-zero. This solution corresponds to a double monolayer with 
a non-zero inter-layer  condensate and a vanishing intra-layer condensate.
\item\label{red} An unconnected solution with zero force between the branes, with $f=0$ and z(r) constant functions for both the D5 brane and the anti-D5 brane, 
but where the branes pinch off before reaching the Poincar\'e horizon.  An embedding of a single D brane which does not reach the 
Poincae\'e horizon is called  a ``Minkowski embedding''.  Since $\psi(r)$ must be $r$-dependent, its asymptotic behaviour is given in \eqref{bcs} with a non-vanishing $c_2$.  This embedding corresponds to a double monolayer with a non-zero intra-layer condensate and a vanishing inter-layer condensate. 
\item A connected $r$-dependent solution, where both $z(r)$ and $\psi(r)$ are nontrivial functions of $r$.  This solution
corresponds to the double monolayer with both an intra-layer and an inter-layer condensate.
\end{enumerate}
This classification of the solutions is summarized in table \ref{typessol}.

\begin{table}[!ht]
	\begin{center}
	{\renewcommand{\arraystretch}{1.1}
		\setlength{\extrarowheight}{2pt}
		\begin{tabular}{c|c|c|}
			\cline{2-3}
			& $f=0$ & $f\neq 0$ \\
			\hline
			\multicolumn{1}{|c|}{\multirow{3}{*}{$c_2=0$}} & \textbf{Type 1} & \textbf{Type 2}\\
		 	\multicolumn{1}{|c|}{}	& unconnected, $\psi=\pi/2$ & connected, $\psi=\pi/2$\\
		  \multicolumn{1}{|c|}{}& BH, chiral symm.  &  inter \\
			\hline
			\multicolumn{1}{|c|}{\multirow{3}{*}{$c_2 \neq 0$}} & \textbf{Type 3} & \textbf{Type 4}\\
			\multicolumn{1}{|c|}{}& unconnected, $r$-dependent $\psi$ & connected, $r$-dependent $\psi$ \\
			\multicolumn{1}{|c|}{}& Mink, intra &  intra/inter\\
				\hline
		\end{tabular}}
		\end{center}	
	\caption{Types of possible solutions, where Mink stands for Minkowski embeddings and BH for black hole embeddings.}
	\label{typessol}
\end{table}

For type 2 and 4 solutions the D5 and the anti-D5 world-volumes
have to join smoothly at a finite $r=r_0$. 
 For these solution the charge density on the brane and on the anti-brane, as well as the value of the constant $f$ that gives the interaction between the brane and the anti-brane, are equal and opposite.

Consider now the solutions of the type 3, types 1 and 2 are just $\psi=\pi/2$. 

The equation  for $\psi$ \eqref{qzero} with $f=0$ simplifies further to
\begin{equation}\label{fzeroqzero}
\frac{r \psi ''+ \psi'}{1+r^2 \psi'^2}-\frac{\psi' r \left(1+3 r^4\right) \sin ^4\psi-2 
   \left(1+r^4\right) \sin ^3\psi \cos \psi }{r \left(1+r^4\right) \sin ^4\psi}=0
\end{equation}
In this case it is obvious from \eqref{zp} that $z(r)$ is a constant. 
Solutions of type 3 are those for which $\psi(r)$ goes to zero at a finite value of $r$, $r_{\rm min}$, so that the two-sphere in the world-volume of the D5 brane shrinks to zero at $r_{\rm min}$.

A solution to \eqref{fzeroqzero}  of this type can be obtained by a shooting technique. The differential equation can be solved from either direction: from $r_{\rm min}$ or from the boundary at $r=\infty$.  In either case, there is a one-parameter family of solutions, from $r_{\rm min}$ the parameter is $r_{\rm min}$, from infinity it is the value of the modulus $c_2$ in \eqref{bcs}, which can be used to impose the boundary conditions at $r\to\infty$ with $c_1=0$. The parameters at the origin $r_{\rm min}$ and at infinity can be varied to find the unique solution that interpolates between the Poincar\'e horizon and the boundary at $r=\infty$.

Consider now the solution  of equation \eqref{qzero} of type 4. In this case we look for a D5 that joins at some given $r_0$  the corresponding anti-D5. At $r_0$, $z'(r_0)\to\infty$ and $r_0$ can be determined by imposing this condition, that, from \eqref{zp} with $q=0$ and $b$ scaled out, reads
 \begin{equation}\label{r0q0}
f^2-r_0^4 \left(1+r_0^4\right) \sin ^4\psi(r_0)=0
\end{equation}
which yields
\begin{equation}\label{bc0}
\psi(r_0)=\sin ^{-1}\left(\sqrt[4]{\frac{f^2}{r_0^4 \left(1+r_0^4\right)}}\right)
\end{equation}
The lowest possible value of $r_0$ is obtained when $\psi=\dfrac{\pi}{2}$ and is given by
\[
r_{0,\text{min}}(f)=\frac{\sqrt[4]{\sqrt{1+4 f^2}-1}}{\sqrt[4]{2}}
\]
Note that $r_{0,\text{min}}$ grows when $f$ grows.

Using \eqref{bc0} we can derive from the equation of motion \eqref{qzero} the condition on $\psi'(r_0)$, it reads 
\begin{equation}\label{bcp0}
\psi'(r_0)=\frac{\left(r_0^4+1\right) \sqrt{\frac{\sqrt{r_0^4+1}}{f}-\frac{1}{r_0^2}}}{2
   r_0^4+1}
\end{equation}
To find the solution let us fix some $\bar{r}$ between $r_0$ and $r=\infty$.  Start with shooting from the origin with boundary conditions \eqref{bc0} and \eqref{bcp0}. \eqref{bc0} leads to $z'(r_0)\to\infty$, but for a generic choice of $r_0$ the solution for $\psi(r)$  does not encounter the solution coming from infinity that has $c_1=0$, we then need to vary the two parameters $r_0$ and $c_2$ in such a way that the two solutions, coming from $r_0$ and from $r=\infty$ meet at some intermediate point.

For $\psi$ and $\psi'$ given by \eqref{bc0} and \eqref{bcp0} at the origin, integrate the solution outwards to $\bar{r}$ and compute $\psi$ and its derivative at $\bar{r}$.  For each solution, put a point on a plot of $\psi'(\bar{r})$ vs $\psi(\bar{r})$, then do the same thing starting from the boundary, $r=\infty$, and varying the coefficient $c_2$ of the expansion around infinity. Where the two curves intersect the $r$-dependent solutions from the two sides match and give the values of the moduli for which there is a solution.

\subsection{Separation and free energy }

\begin{figure}[!ht]
\centerline{\includegraphics[scale=1]{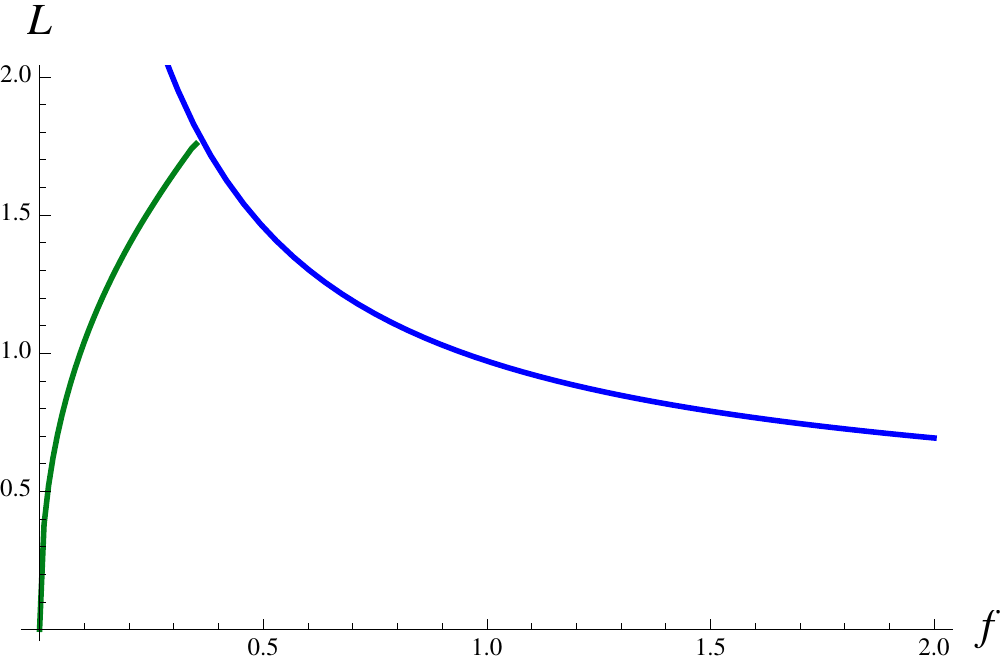}}
\caption{\small The   separation of the monolayers, $L$,  is plotted on the vertical axis and the force parameter $f$ is plotted on
the horizontal axis.  The branch indicated by the blue line is for the   
constant connected (type 2) solution.  (It is a graph of equation (\ref{Lanalytic}).)  The green line is for the $r$-dependent connected (type 4) solution. }
\label{fig:branesepq=0}
\end{figure}

There are then four types of solutions of the equation of motion \eqref{qzero} representing double monolayers with a magnetic field, of type 1, 2, 3 and 4. 
Solutions 1 and 3 are identical to two independent copies of a single mono-layer with $B$ field solution, sitting at a separation $L$. 
The brane separation for the solutions of type 2 and 4 is given in \eqref{L} (for $q=0$ in this case) and it is plotted in fig.~\ref{fig:branesepq=0}, the blue line gives the analytic curve \eqref{Lanalytic}, keeping into account that also $r_0$ is a function of $f$ through equation \eqref{r0}. For the $r$-dependent solution, green line, instead, $r_0$ is defined as a function of $f$ by equation \eqref{r0q0}, once the solution $\psi(r)$ is known numerically.

We shall now compare the free energies of these solutions as a function of the separation to see at which separation one becomes preferred with respect to the other. 

\begin{figure}[!ht]
\centerline{\includegraphics[scale=1]{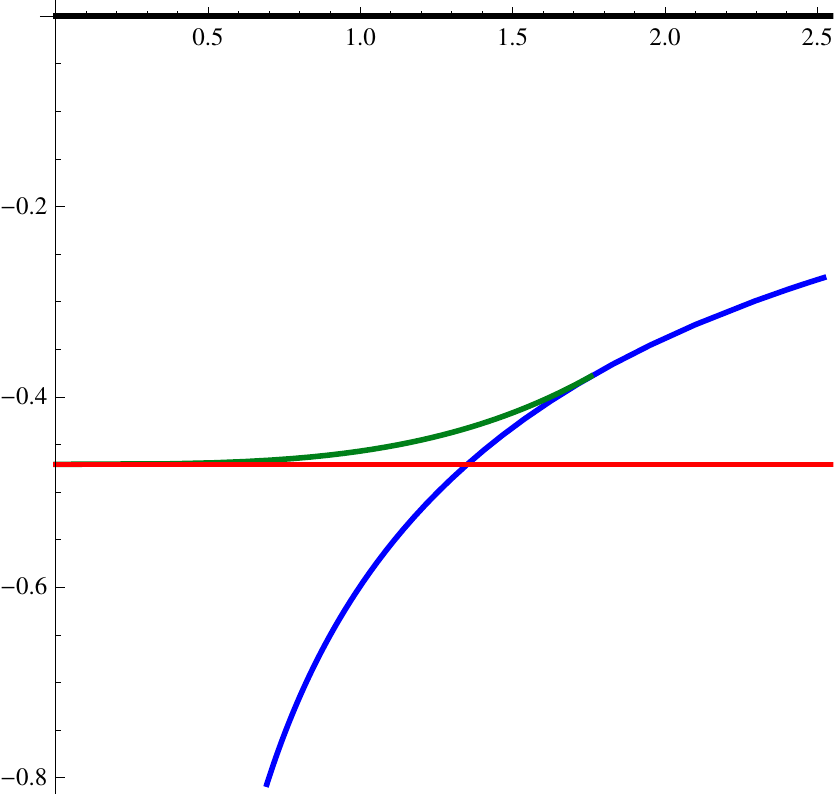}}
\begin{picture}(0,0)(0,0)
\put(340,238){$L$}
\put(103,253){$\Delta \CF_1$}
\end{picture}
\caption{\small Double monolayer in a magnetic field, where each monolayer is charge neutral.  The 
regularized free energy $\Delta \CF_1$  is plotted on the 
vertical axis, 
and the inter-layer separation $L$ (in units of $1/\sqrt{b}$), which is plotted on the horizontal axis.  The blue line corresponds to
the connected solution (type 2), the red line to the unconnected solution (type 3) and the green line to the connected $r$-dependent solutions (type 4). 
All solutions are regulated by subtracting the free energy of  the constant unconnected solution of type 1. The latter is the black line at the top
of the diagram.  The type 1 and type 4 solutions exist but they never have the lowest energy.  For large $L$, the type 3 solution is preferred and small $L$ 
the type 2 solution is more stable. This reproduces results quoted in reference  \cite{Evans:2013jma}. }
\label{fig:FEdq0}
\end{figure}
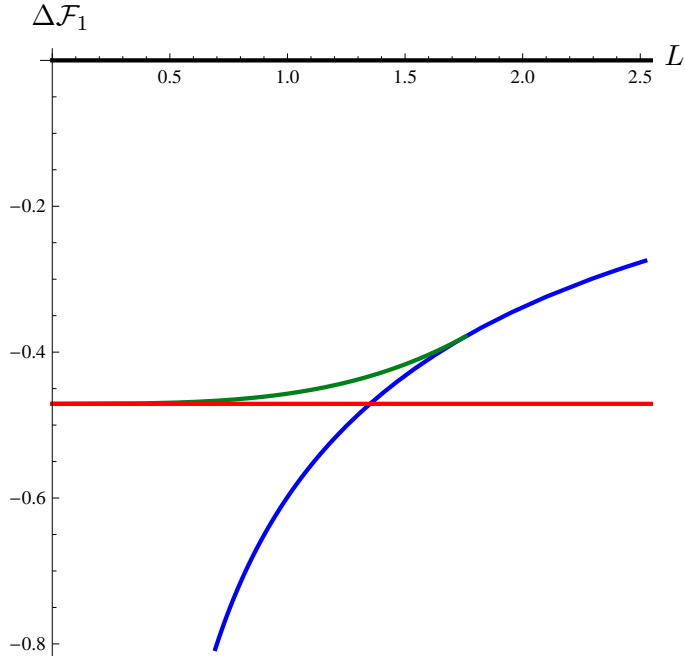

Since we want to compare solutions at fixed values of $L$ the correct quantity that provides the free energy for each configuration is given by the action evaluated on the corresponding solution
\begin{equation}
	\CF_1[L] = \int_{r_0}^\infty \,dr\,  \left(1+r^4\right) \sin ^4\psi  \frac{r^2 \sqrt{ 1+r^2 \psi '^2}}{\sqrt{
   r^4 \left(1+r^4\right) \sin ^4\psi-f^2}} \, .
\end{equation}
Note that this formula is obtained from \eqref{SLmu}, by setting $q=0$ and performing the rescaling \eqref{rescbq}. The dependence of $\CF_1$ on $L$ is
implicit (recall that we can in principle trade $f$ for $L$).
This free energy is divergent since in the large $r$ limit the argument of the integral goes as $\sim r^2$.
However, in order to find the energetically favored configuration,
we are only interested in the difference between the free energies of two solutions, which is always finite.
We then choose the free energy of the unconnected ($f=0$) constant $\psi=\pi/2$ solution, type 1,
as the reference free energy (zero level), so that any other 
(finite) free energy can be defined as 
\begin{align}
&\Delta \CF_1(\psi;f)= \CF_1(\psi;f)-\CF_1(\psi=\pi/2;f=0)=
\nonumber\\
	& \int_{r_0}^\infty \,dr\, \left( \left(1+r^4\right) \sin ^4\psi  \frac{r^2 \sqrt{ 1+r^2 \psi '^2}}{\sqrt{
   r^4 \left(1+r^4\right) \sin ^4\psi-f^2}}-
   \sqrt{ 1+r^4}\right)
   \nonumber\\ 
   & - {r_0} \, _2F_1\left(-\frac{1}{2},\frac{1}{4};\frac{5}{4};-{r_0}^4\right)\, .
\end{align}
where the last term is a constant that keeps into account that the $\psi=\pi/2$ disconnected solution reaches the Poincar\'e horizon, whereas the other solutions do not.
It turns out that, in this particular case where the D5 brane and the anti-D5 brane are both charge neutral, 
the solutions of type 1 and 4 always have a higher free energy than solutions 2 and 3.
By means of numerical computations we obtain for the free energy  $\Delta \CF_1$ of the solutions 2, 3 and 4,
the behaviours depicted in Figure \ref{fig:FEdq0}.  This shows that the 
dominant configuration is the connected one with an inter-layer condensate for small brane separation $L$ and the disconnected
one, with only an intra-layer condensate, for large $L$. The first order transition between the two phases takes place at $L\simeq 1.357$ in agreement with 
the value quoted in reference \cite{Evans:2013jma}.
 
\vspace*{1em}

\section{Double monolayer with a magnetic field and a charge-balanced chemical potential}

We shall now study the possible configurations for the D5-anti D5 probe branes in the $AdS_5\times S^5$
background, with a magnetic field and a chemical potential. The chemical potentials are balanced in such a way
that the chemical potential on one monolayer induces a density of electrons and the chemical potential on the other monolayer
induces a density of holes which has identical magnitude to the density of electrons.    
Moreover, the chemical potentials are exactly balanced so that the density of electrons and the density
of holes in the respective monolayers are exactly equal.   Due to the particle-hole symmetry of the quantum field theory, it is sufficient
that the chemical potentials have identical magnitudes.    The parameters that we keep
fixed in our analysis are the magnetic field $b$, the monolayer separation $L$ and the chemical potential $\mu$.

In order to derive the allowed configurations we have to solve equation~\eqref{eompsi} for 
$\psi$ as well as equation~\eqref{a0p} for the gauge potential $a_0$ and equation~\eqref{zp} for $z$.
In practice the difficult part is to find all the solutions of the equation of motion for $\psi$, which is 
a non-linear ordinary differential equation. Once one has a solution for $\psi$ it is straightforward to 
build the corresponding solutions for $z$ and $a_0$, simply by plugging the solution for $\psi$ into the 
equations \eqref{a0p}-\eqref{zp} and integrating them.

It should be noted that any solution of the equation \eqref{a0p} for the gauge potential   $a_0(r)$ always has an ambiguity in that  $a_0(r)+$constant is
also a solution.  The constant is fixed by remembering that $a_0(r)$ is the time component of a vector field and it should therefore vanish at the 
Poincar\'e horizon.     When the charge goes to zero, $a_0=$constant is the only solution of  equation \eqref{a0p} and this condition puts the constant to 
zero.  Of course, this is in line with particle-hole symmetry which tells us that the state with chemical potential set equal to zero 
has equal numbers
of particles and holes.   The results of the previous section, where $\mu$ and $q$ were equal to zero,  
care a special case of what we will derive below. 

\subsection{Solutions for $q\neq 0$}

Now we consider the configurations with a charge density different from zero.
The differential equation for $\psi$ in this case is 
\begin{equation}\label{eomq}
\frac{r \psi ''+ \psi '}{1+r^2 \psi '^2}-\frac{\psi ' \left(f^2+q^2r^4+r^4 \left(1+3 r^4\right) \sin ^4\psi\right)-2 r^3
   \left(1+r^4\right) \sin ^3\psi \cos \psi }{f^2-q^2r^4-r^4 \left(1+r^4\right) \sin ^4\psi}=0\, .
\end{equation}

As usual, we shall look for solutions with $c_1=0$ in equation \eqref{bcs}.  We can again distinguish   four types of solutions according
to the classification of table \ref{typessolq}.  The main difference between the solutions summarized in  table \ref{typessolq}
and those in table 1 are that the type 3 solution now has a black hole, rather than a Minkowski embedding.  This is a result of
the fact that, as explained in section 1,  the world-volume of a D5 brane that carries electric charge density must necessarily 
reach the Poincar\'e horizon if it does not join with the anti-D5 brane. The latter, where it reaches the Poincar\'e horizon,  is an 
un-gapped state and it must be so even when there is an intra-layer exciton condensate.
It is, however, incompatible with an inter-layer condensate.

\begin{table}[!ht]
	\begin{center}
	{\renewcommand{\arraystretch}{1.1}
		\setlength{\extrarowheight}{2pt}
		\begin{tabular}{c|c|c|}
			\cline{2-3}
			& $f=0$ & $f\neq 0$ \\
			\hline
			\multicolumn{1}{|c|}{\multirow{3}{*}{$c_2=0$}} & \textbf{Type 1} & \textbf{Type 2}\\
		 	\multicolumn{1}{|c|}{}	& unconnected, $\psi=\pi/2$ & connected, $\psi=\pi/2$\\
		  \multicolumn{1}{|c|}{}& BH, chiral symm.  & inter \\
			\hline
			\multicolumn{1}{|c|}{\multirow{3}{*}{$c_2\neq 0$}} & \textbf{Type 3} & \textbf{Type 4}\\
			\multicolumn{1}{|c|}{}& unconnected, $r$-dependent $\psi$ & connected, $r$-dependent $\psi$ \\
			\multicolumn{1}{|c|}{}& BH, intra & intra/inter\\
				\hline
		\end{tabular}}
		\end{center}	
	\caption{Types of possible solutions for $q\neq 0$.}
	\label{typessolq}
\end{table}

Type 1 solutions are trivial both in $\psi$ and $z$ (they are both constants).
They correspond to two parallel black hole (BH) embeddings for the 
D5 and the anti-D5. This configuration is the chiral symmetric one.
In type 2 solutions the chiral symmetry is broken by the inter-layer condensate ($f\neq 0$): 
In this case the branes have non flat profiles in the $z$ direction.
Solutions of type 3 and 4 are $r$-dependent and consequently are the really non-trivial ones to find. 
Type 3 solutions have non-zero expectation value of the intra-layer condensate
and they can be only black hole embeddings, this is the most significant difference with the zero charge case. 
Type 4 solutions break chiral symmetry in both the inter- and intra-layer channel.
For type 2 and 4 solutions the D5 and the anti-D5 world-volumes
have to join smoothly at a finite $r=r_0$.

Now we look for the non-trivial solutions of equation~\eqref{eomq}. We start considering the solutions of type 4. 
We can build such a solution requiring that the D5 profile smoothly joins at some given $r_0$  the corresponding anti-D5 profile. 
The condition that has to be satisfied in order to have a smooth solution for the connected D5/anti-D5 world-volumes is $z'(r_0)\to\infty$ which, 
from \eqref{zp} (with $b$ scaled to 1), corresponds to the condition
\begin{equation}\label{r0q}
f^2 -r_0^4 \left[q^2+\left(1+r_0^4\right) \sin ^4\psi(r_0)\right]=0\, .
\end{equation}
From this we can determine the boundary value $\psi(r_0)$
\begin{equation}\label{bcq}
\psi(r_0)=\arcsin \left(\sqrt[4]{\frac{f^2-q^2 r_0^4}{r_0^4 \left(1+r_0^4\right)}}\right)\, .
\end{equation}
Note that the request that $0\le \sin\psi(r_0)\le 1$ fixes  both a lower and an upper bound on $r_0$
\[
r_{0,\text{min}}(f,q)=\frac{\sqrt[4]{\sqrt{(1+q^2)^2+4 f^2}-1-q^2}}{\sqrt[4]{2}}\, , \qquad r_{0,\text{max}}(f,q)=\sqrt{\frac{f}{q}}\, .
\]
Using \eqref{bc0} we can derive from the equation of motion \eqref{eomq} the condition on $\psi'(r_0)$, which reads
\begin{equation}\label{bcpq}
\psi'(r_0)=\frac{\left(r_0^4+1\right)\left(f^2-q^2r_0^4\right) \sqrt{\sqrt{\frac{r_0^4+1}{f^2-q^2r_0^2}}-\frac{1}{r_0^2}}}{f^2\left(2
   r_0^4+1\right)-q^2r_0^8}
\end{equation}
We can then build such solutions imposing the conditions \eqref{bcq} and \eqref{bcpq} at $r_0$, where $r_0$ is the modulus.
With the usual shooting technique we then look for solutions that also have the desired behavior at
infinity, \textit{i.e.} those that match the boundary conditions \eqref{bcs} with $c_1=0$.
It turns out that in the presence of a charge density there are solutions of type 4 
for any value of $q$, in this case, however, these solutions will play an important role in the phase diagram.

Next we consider the solutions of type 3. These solutions can be in principle either BH or Minkowski embeddings. 
However when there is a charge density different from zero only BH embeddings are allowed. A charge density on the
D5 world-volume is indeed provided by fundamental strings stretched between the D5 and the Poincar\'e horizon. These strings
have a tension that is always greater than the D5 brane tension and thus they pull the D5 down to the Poincar\'e horizon~\cite{Kobayashi:2006sb}.
For this reason when $q\neq 0$ the only disconnected solutions we will look for  are BH embedding. 
Solutions of this kind with $c_2\neq 0$ can be built numerically along the lines of ref. \cite{Grignani:2012jh}.
Note that because of the
equation of motion they must necessarily have $\psi(0)=0$\footnote{Actually also the condition $\psi(0)=\pi/2$ is allowed by the equation
of motion, but this would correspond to the constant solution with $c_2=0$, namely the type 1 solution.}.

\subsection{Separation and free energy}

\begin{figure}[!ht]
\centerline{\includegraphics[scale=1]{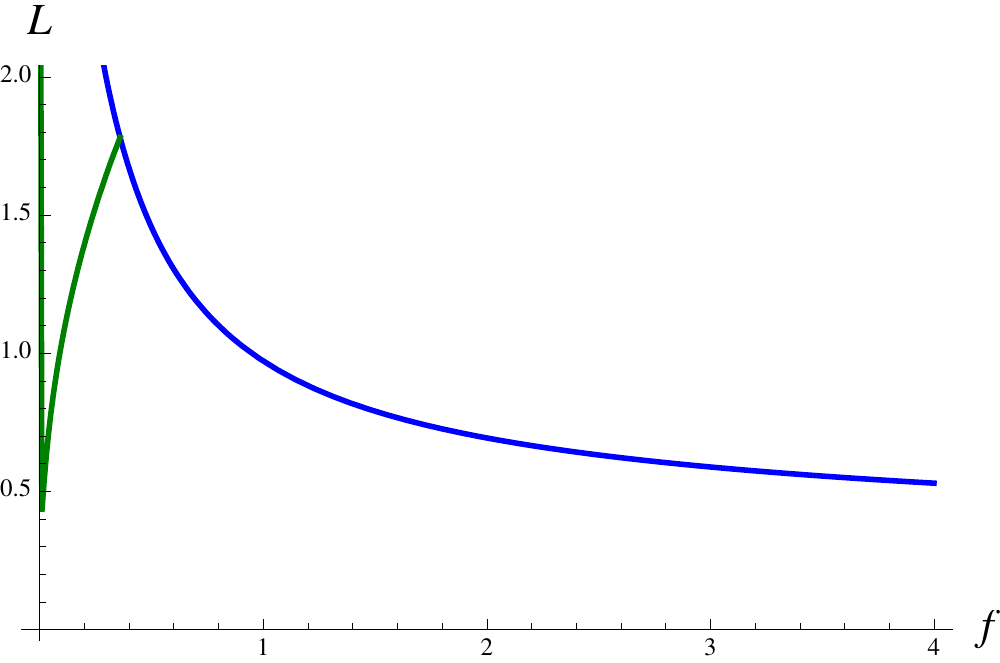}}
\begin{picture}(0,0)(0,0)
\end{picture}
\caption{\small The   separation of the monolayers, $L$,  is plotted on the vertical axis and the force parameter $f$ is plotted on
the horizontal axis, in the case where the monolayers have charge densities and $q=0.01$.  The branch indicated by the blue line is for the   
constant connected (type 2) solution.  The green line is for the $r$-dependent connected (type 4) solution. }
\label{fig:branesepq}
\end{figure}

The brane separation is given in \eqref{L}  and for the solutions of type 2 and 4 is plotted in fig.~\ref{fig:branesepq}  for $q=0.01$, the blue line gives the curve \eqref{Lanalytic}, keeping into account that also $r_0$ is a function of $f$ through equation \eqref{r0}. For the $r$-dependent solution instead $r_0$ is defined as a function of $f$ by equation \eqref{r0q}, once the solution $\psi(r)$ is known numerically.
The $r$-dependent connected solution, green line, has two branches one in which $L$ decreases with increasing $f$ and the other one in which $L$ increases as $f$ increases. It is clear from the picture that when $q\to 0$ one of the branches of the green solution disappears and  fig.~\ref{fig:branesepq}  will become  identical to fig.~\ref{fig:branesepq=0}.

Once we have determined all the possible solutions, it is necessary to study which configuration is 
energetically favored. We shall compare the free energy of the solutions at fixed values of $L$ and $\mu$, since
this is the most natural experimental condition for the double monolayer system. 
Thus the right quantity to define the free energy is the action
\eqref{SLmu}, which after the rescaling \eqref{rescbq} is given by
\begin{equation}\label{EnLmu}
\CF_1[L,\mu] = \int_{r_0}^\infty \,dr\,  \left(1+r^4\right) \sin ^4\psi  \frac{r^2 \sqrt{ 1+r^2 \psi '^2}}{\sqrt{
   r^4 \left(1+r^4\right) \sin ^4\psi+q^2 r^4-f^2}}
\end{equation}
As usual we regularize the divergence in the free energy by considering the difference of free energies of pairs of solutions, 
which is really what we are interested in. So we define a regularized free energy $\Delta \CF_1$ by subtracting 
to each free energy that of the unconnected ($f=0$) constant $\psi=\pi/2$ solution, 
\begin{equation}\label{defDEq}
	\Delta \CF_1(\psi;f,q)\equiv \CF_1(\psi;f,q)-\CF_1(\psi=\pi/2;f=0,\hat{q}). 
\end{equation}

As we already noticed, the free energy \eqref{EnLmu} and consequently $\Delta \CF_1$
are implicit functions of $L$ and $\mu$, via $f$ and $q$, which are the parameters that we really have under control in the 
calculations. Thus when computing the regularized free energy $\Delta \CF_1$ we have to make sure that 
the two solutions involved have the same chemical potential.\footnote{In principle we also have to make sure that the two solutions
have the same $L$. However this is not necessary in practice, since in $\Delta \CF_1$ we use as reference free energy that of an unconnected solution,
which therefore is completely degenerate in $L$. Indeed the unconnected configuration is given just by two copies of the
single D5 brane solution with zero force between them, which can then be placed at any distance $L$.}
This is the reason why in the definition  of $\Delta \CF_1$ \eqref{defDEq} we subtract the free energy of two solutions with different values of $q$:
the $\hat{q}$ in \eqref{defDEq} is in fact the value of the charge such that the chemical potential of the regulating solution $(\psi=\pi/2;f=0)$
equals that of the solution we are considering $(\psi;f,q)$.
To be more specific, for a solution with chemical potential $\mu$, which can be computed numerically through \eqref{mu} (or through \eqref{mupi2}
for the $\psi=\pi/2$ case),  $\hat{q}$ must satisfy 
\[
	\mu(\psi=\pi/2;f=0,\hat{q})\equiv\frac{4 \Gamma\left(\frac{5}{4}\right)^2\hat{q}}{\sqrt{\pi }(1+\hat{q}^2)^{1/4}}=\mu
\]
and therefore it is given by
\[
	\hat{q}=\sqrt{\frac{\CJ+\sqrt{\CJ(\CJ+4)}}{2}}\, , \qquad \CJ\equiv \frac{\pi^2 \mu^4}{\left(2 \Gamma\left(\frac{5}{4}\right)\right)^8}\, .
\]

For the type 2 solution the regularized free energy density can be computed analytically. Reintroducing back the magnetic filed, it reads
\begin{equation}
\begin{split}\label{analyticenergytotal}
\Delta \CF_1[L,\mu]=&\int_{r_0}^\infty\, dr\left(b^2+r^4\right) \left(\frac{r^2 r_0^2}{ \sqrt{\left(r^4-r_0^4\right) \left(f^2+r^4
   r_0^4\right)}}-\frac{1}{\sqrt{b^2+\hat q^2+r^4}}\right) \\
   &-\int^{r_0}_0\, dr\frac{b^2+r^4}{\sqrt{b^2+\hat q^2+r^4}}\\
	&=\frac{\sqrt{\pi } \Gamma \left(-\frac{3}{4}\right) }{16 r_0 \sqrt[4]{b^2+\hat{q}^2} \Gamma \left(\frac{3}{4}\right)}
	\left[\sqrt[4]{b^2+\hat{q}^2} \left(2 r_0^4 \,
   _2F_1\left(-\frac{3}{4},\frac{1}{2};\frac{3}{4};-\frac{f^2}{r_0^8}\right)\right. \right.\\
   &\left. \left. -3  \left(b^2+r_0^4\right) \,
   _2F_1\left(\frac{1}{4},\frac{1}{2};\frac{3}{4};-\frac{f^2}{r_0^8}\right)\right)+\sqrt{2}
   r_0 \left(2 b^2-\hat{q}^2\right)\right]
   \end{split}
\end{equation}
where $r_0$ is given in \eqref{r0}.

\begin{figure}[!ht]
\begin{minipage}{.5\textwidth}
\vspace{5pt}
\includegraphics[scale=.8]{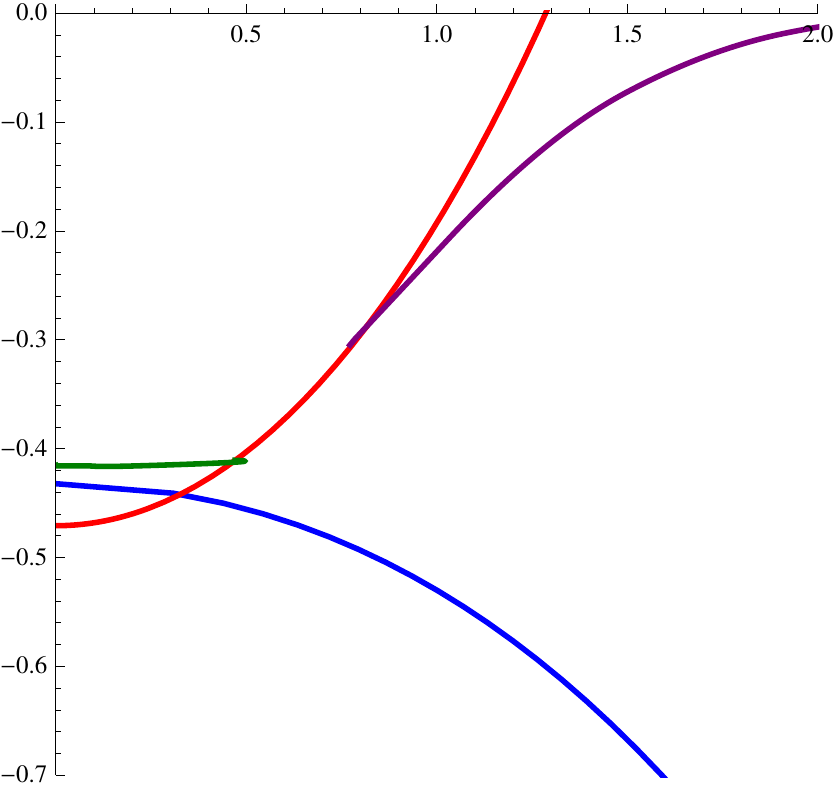}
		\begin{picture}(0,0)(0,0)
			\put(-191,184){$\Delta \CF_1$}
			\put(1,176){$\mu$}
			\put(-80,80) {$L=1.5$}
		\end{picture}
	\end{minipage}
\begin{minipage}{.49\textwidth}
		\includegraphics[scale=.8]{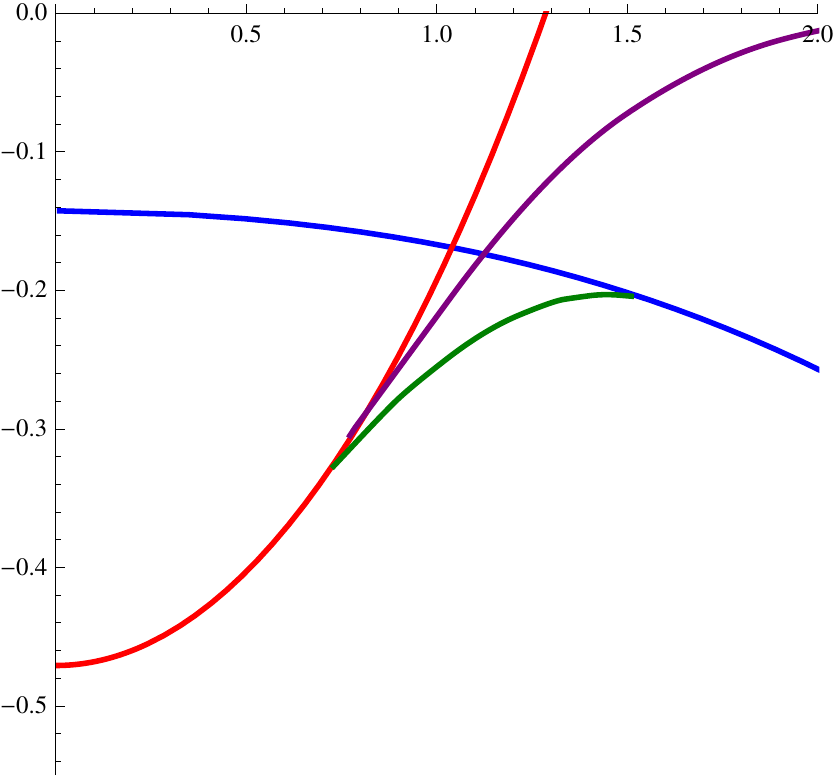}
		\begin{picture}(0,0)(0,0)
			\put(-191,182){$\Delta \CF_1$}
			\put(1,174){$\mu$}
			\put(-80,70) {$L=5$}
		\end{picture}
	\end{minipage}	
	\caption{Plots of the free energies as a function of the chemical potential: type 2 (blue line), type 3 (red-line) and type 4 (green line) solutions for $L=1.5$ and 
		$L=5$.}
		\label{fig:Energies234}
\end{figure}
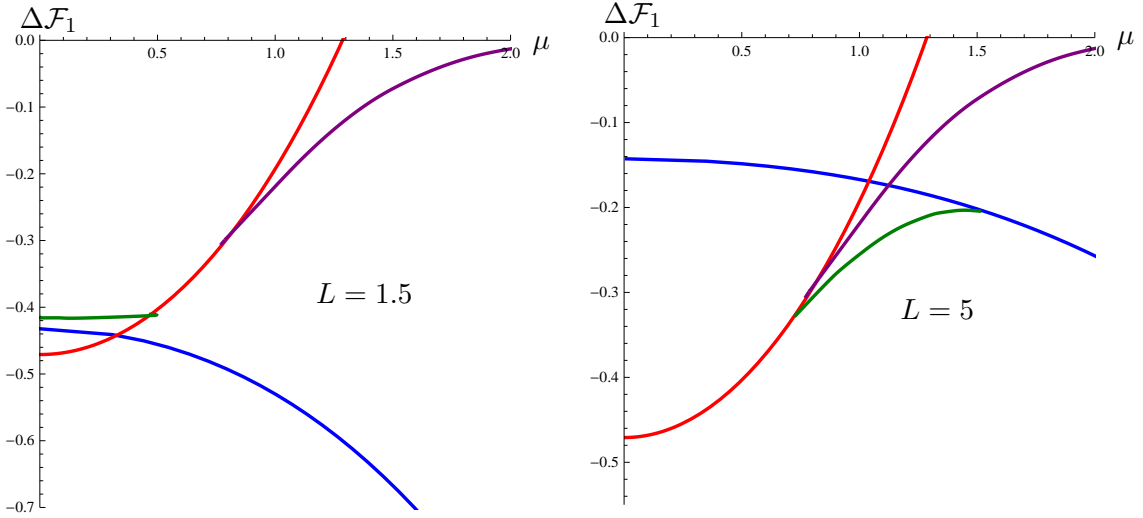

A comparison of the free energies of the various solutions for $L=1.5$ and $L=5$ is given in figure \ref{fig:Energies234}. The chirally symmetric solution would be along the $\mu$-axis since it is the solution we used to regularize all the free energies. It always has a higher free energy, consequently, the chirally symmetric phase is always metastable.

\subsection{Phase diagrams}

\begin{figure}[!ht]
	\begin{center}
	\includegraphics[scale=1.25]{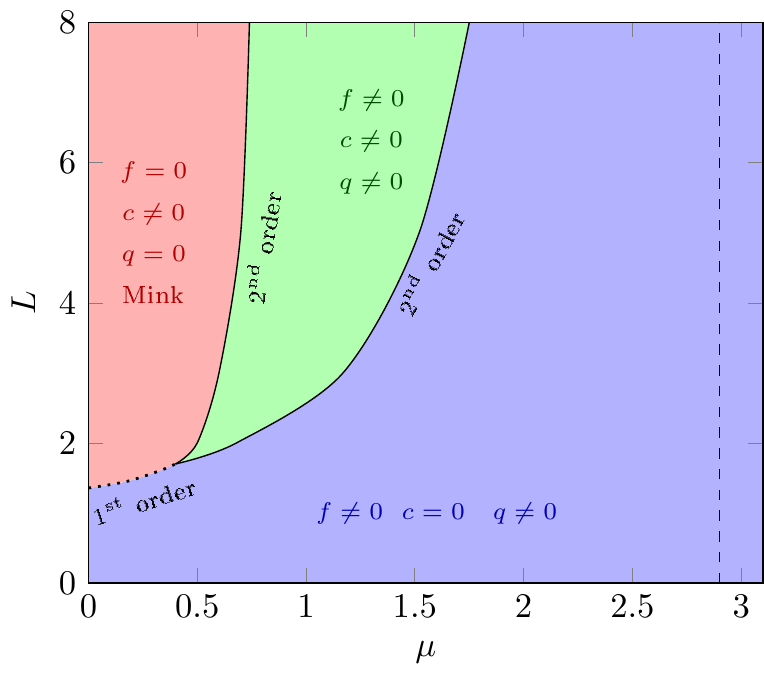}
	\caption{Phase diagram of the D3-probe-D5 branes system with balanced charge densities.  Layer separation is plotted on the vertical axis
	and chemical potential $\mu$ for electrons in one monolayer and holes in the other monolayer is plotted
	on the horizontal axis. The units are the same as in figure \ref{fig:phased0}. }
	\label{fig:phased}	
\end{center}
\end{figure}

Working on a series of constant $L$ slices we are then able to draw the phase diagram $(\mu,L)$ for the system.  For the reader convenience we reproduce the phase diagram that we showed in the introduction in fig.~\ref{fig:phased} (here the labels are rescaled however).
We see that the dominant phases are three: 
\begin{itemize}
	\item The connected configuration with $c_2=0$ (type 2 solution) where the flavor
	symmetry is broken by the inter-layer condensate (blue area); 
	\item The connected  configuration with $c_2\neq 0$ and $f\neq 0$
	(type 4 solution with $q\neq 0$) where the chiral symmetry is broken by the intra-layer condensate and the flavor symmetry is broken by the inter-layer condensates (green area);
	\item The unconnected Minkowski embedding configuration with $c_2\neq 0$ 
	(type 3 solution with $q=0$) where the chiral symmetry is broken by the intra-layer condensate (red area);
\end{itemize}
Note that in all these three phases chiral symmetry is broken.

\begin{figure}
\begin{center}
	\includegraphics[scale=1.2]{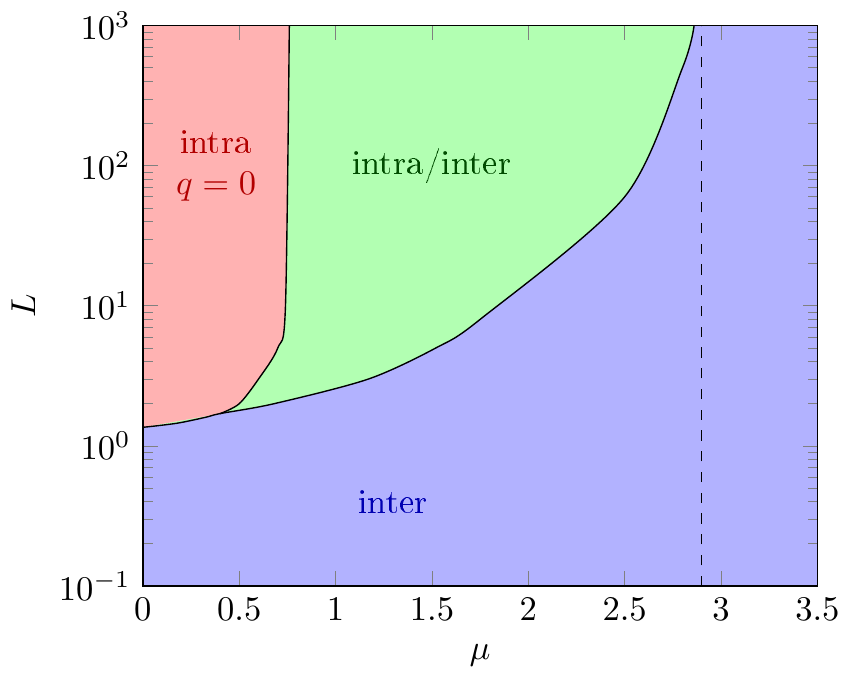}
	\end{center}
\caption{Phase diagram for large separation between the layers.}
	\label{fig:phasedlargeL}
\end{figure}

As expected, for small enough  $L$ the connected configuration is the dominant one. 
We note that for $L\lesssim 1.357$, which, as we already pointed out,
is the critical value for $L$ in the zero-chemical potential case, the connected configuration is always preferred for any value of
$\mu$. 
When $1.357\lesssim L \lesssim 1.7$  the system faces a second order phase transition from
the unconnected Minkowski embedding phase -- favored for small values of $\mu$ -- 
to the connected phase --
favored at higher values of $\mu$.
When $L \gtrsim 1.7$ as the chemical potential varies the system undergoes two phase transitions:
The first happens at $\mu\simeq 0.76$ and it is a second order transition from the unconnected Minkowski embedding phase to the
connected phase with both condensates. Increasing further the chemical potential
the system switches to the connected phase with only an intra-layer condensate again via a second order transition.

Therefore it is important to stress that, with a charge density, a phase with coexisting inter-layer and intra-layer condensates can be the energetically
preferred state.  Indeed, it is the energetically favored solution in the green area and corresponds to states of the  
double monolayer with both the inter-layer and intra-layer condensates.

The behavior of the system at large separation between the layers is given in fig.~\ref{fig:phasedlargeL}. For $L \to \infty$ the phase transition 
line between the green and the blue area approaches a vertical asymptote at $\mu\simeq 2.9$. The connected solutions for $L\to\infty$ become the corresponding, $r$-dependent or $r$-independent disconnected solutions. Thus for an infinite distance between the layers
we recover exactly the behavior of a single layer \cite{Evans:2010hi} where at $\mu\simeq 2.9$ the system undergoes a BKT transition between 
the intra-layer BH embedding phase to the chiral symmetric one \cite{Jensen:2010ga}.

\begin{figure}[!ht]
\vskip 2mm
\centerline{\includegraphics[scale=1]{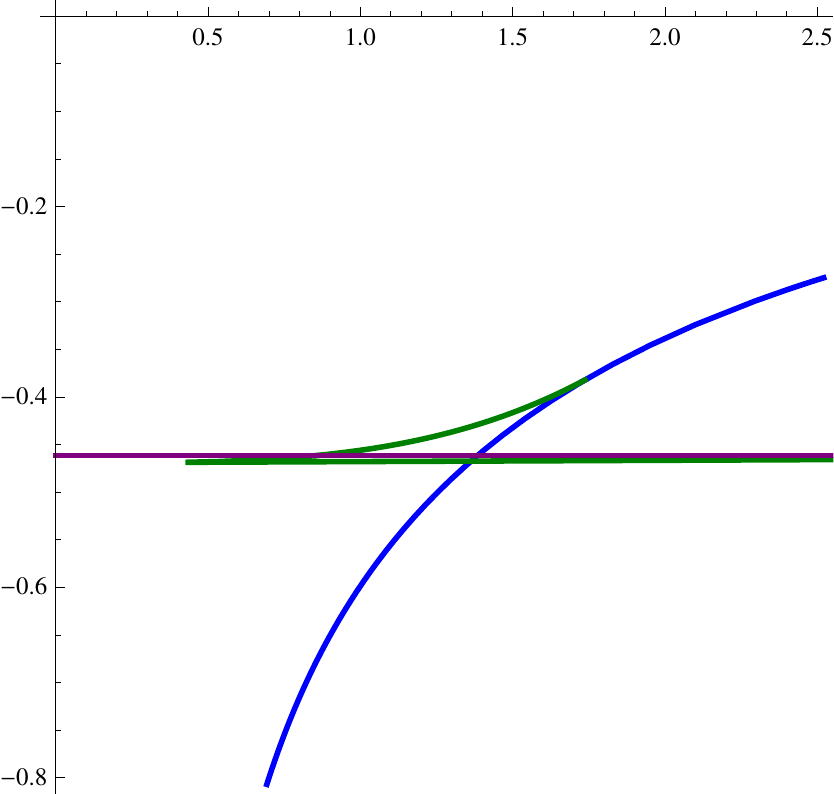}}
\begin{picture}(0,0)(0,0)
\put(340,239){$L$}
\put(103,251){$\Delta \CF_2$}
\end{picture}
\caption{\small Free energy difference $\Delta \CF_2$ as a function of $L$ for $q=0.01$, the two branches of the $r$-dependent connected solution, green line, tend to become just the one of fig.~\ref{fig:FEdq0}.}
\label{fig:FEdq}
\end{figure}
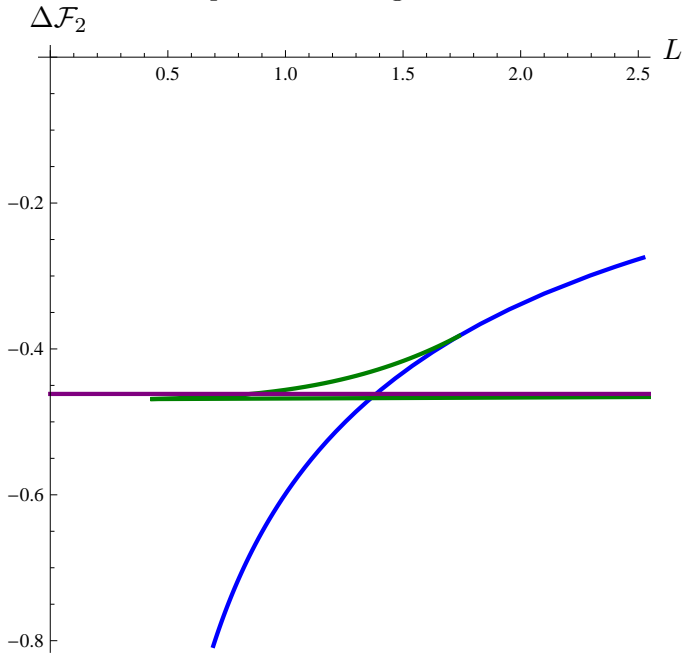

\begin{figure}
	\begin{center}
	\includegraphics[scale=1]{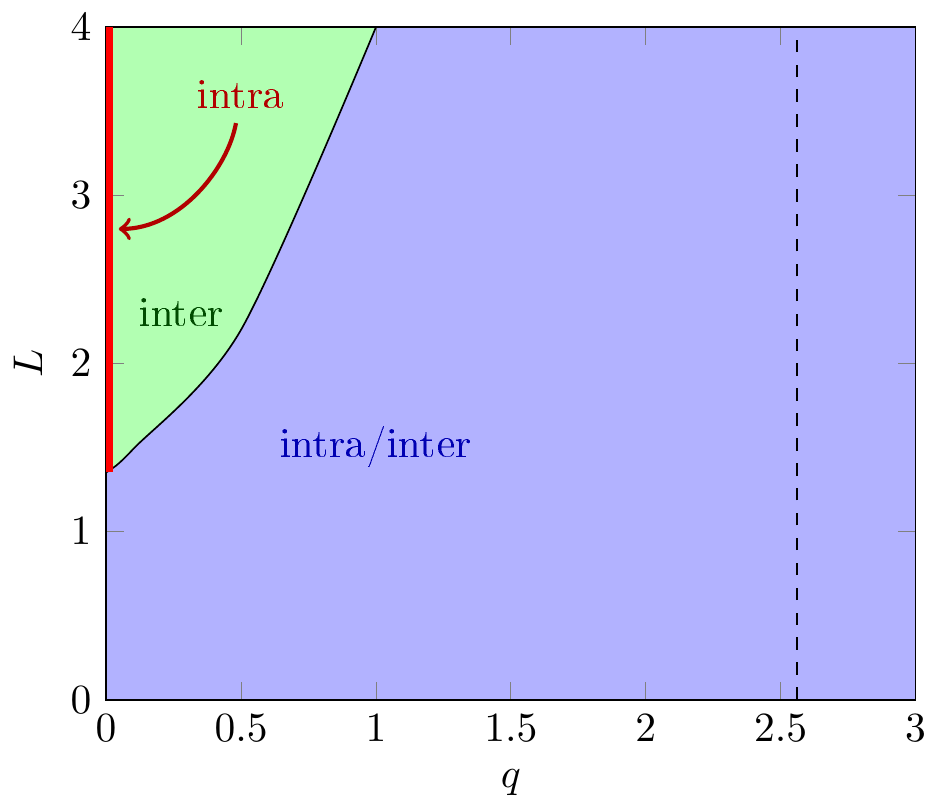}
	\end{center}	
\caption{Phase diagram in terms of the brane separation $L$ and the charge density $q$.}
		\label{fig:phasedLq}
\end{figure}

It is interesting to consider also the phase diagram in terms of the brane separation and charge density fig.~\ref{fig:phasedLq}. In this case the relevant free energy function that has to be considered is the Legendre transformation of the action with respect to $q$, namely the Routhian  $\mathcal{F}_2[L,q]$ defined in equation~\eqref{R1}. For the regularization of the free energy we choose proceed in analogy as before: For each solution of given $q$ and $L$ we subtract the free energy of the constant disconnected (type 1) solution with the same charge $q$, obtaining the following regularized free energy
\begin{equation}\label{DF2}
	\Delta \CF_2[L,q]\equiv \CF_2[L,q]-\CF_2(\psi=\pi/2;f=0)[q]. 
\end{equation}
In fig.~\ref{fig:phasedLq} the phase represented by the red region in fig.~\ref{fig:phased},  is just given by a line along the $q=0$ axis. By computing the explicit form of the free energies as function of the brane separation $L$, it is possible to see in fact that the $r$-dependent connected solution has two branches. These branches reflect the fact that also the separation $L$ has two branches as a function of $f$, as illustrated in fig.~\ref{fig:branesepq}. In the limit $q\to 0$ one of these branches tends to overlap to the $r$-dependent disconnected solution and for $q=0$ disappears. This is illustrated in fig.~\ref{fig:FEdq} where one can see that in the $q\to 0$ limit the free energy difference as a function of $L$ goes back to that represented in fig.~\ref{fig:FEdq0}.

\section{Double monolayers with un-matched charge densities}

We now consider a more general system of two coincident D5 branes and two coincident anti-D5 branes, with total charges $Q=q_1+q_2>0$ and $-\bar{Q}$ where $\bar{Q}=q_3+q_4>0$.
Then, unlike before, this corresponds to a double monolayer with unpaired charge on the two layers.
For such a system we are interested in determining the most favored configuration, \textit{i.e.} to
find out how the charges $Q$ and $\bar{Q}$ distribute among the branes and which types of solutions
give rise to the least free energy for the whole system. Since the parameter that we take under control 
is the charge, and not the chemical potential, we shall use the free energy $\CF_2$, defined in equation~\eqref{F2}, in order to compare the different solutions. 

What we keep fixed in this setup are the overall charges $Q$ and $\bar{Q}$ 
in the two layers, while we let the charge on each brane vary: Namely the $q_i$ vary
with the constraints that  $Q=q_1+q_2$ and $\bar{Q}=q_3+q_4$ are fixed. 
Then we want to compare configurations with different values for the charges $q_i$ on the single branes. For this reason we must choose a regularization of the free energy that does not depend on the charge on the single brane, and clearly the one that we used in the previous section is not suitable. The most simple choice of such a regularization consists in subtracting to the integrand of the free energy only its divergent part in the large $r$ limit, which is $r^2$. We denote this regularized free energy as $\Delta \CF_{2,r}$.

Without loss of generality we suppose that $Q>\bar{Q}$. Then for simplicity we fix the values of the charges to $Q=0.15$ and $\bar{Q}=0.1$ and the separation between the layers to $L=1$. There are two possible cases.

\begin{enumerate}[$(i)$]
\item\label{D5-1}
A configuration in which the D5 brane with charge $q_1$ is described by a black hole embedding whereas the D5 brane with charge $q_2$ is connected with the anti-D5 brane with charge $q_3$, so that $q_2=q_3$. Then we have
\[
q_2=q_3=\bar{Q}-q_4~,~~~~~q_1=Q-q_2=Q-\bar{Q}+q_4
\]

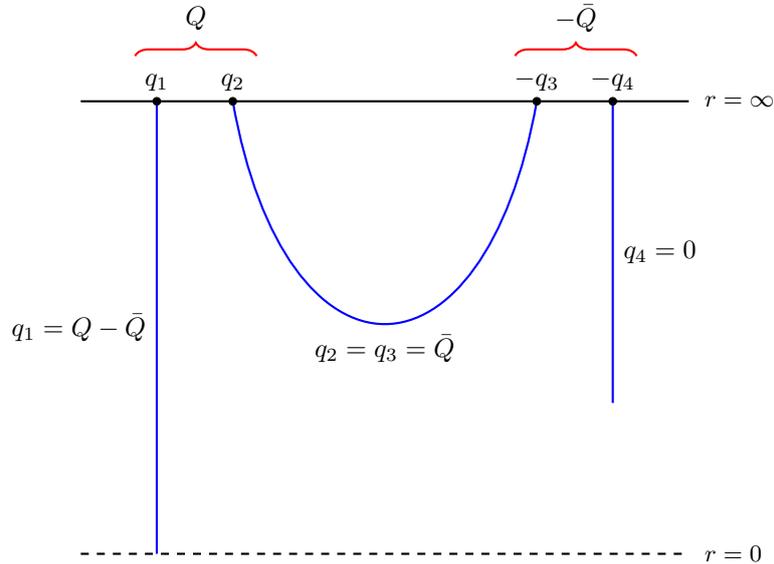
\begin{figure}[h!]
\begin{center}

\begin{tikzpicture}[scale=1]
	\coordinate (q1) at (-3,6);
	\coordinate (q2) at (-2,6);
	\coordinate (q3) at (2,6);
	\coordinate (q4) at (3,6);
	\draw[thick] (-4,6) -- +(8,0) node[right=2pt] {\footnotesize $r=\infty$};
	\draw[thick,dashed] (-4,0) -- +(8,0) node[right=2pt] {\footnotesize $r=0$};
	\draw[thick,blue] (q2) .. controls +(100:-4) and +(80:-4) .. (q3)node [midway,below,black] {\small $q_2=q_3=\bar{Q}$};
	\draw[thick,blue] (q1) -- +(0,-6) node [midway,left,black] {\small $q_1=Q-\bar{Q}$};
	\draw[thick,blue] (q4) -- +(0,-4) node [midway,right,black] {\small $q_4=0$};;
	\filldraw (q1) circle (.05) node[above] {\small $q_1$};
	\filldraw (q2) circle (.05) node[above] {\small $q_2$};
	\filldraw (q3) circle (.05) node[above] {\small $-q_3$};
	\filldraw (q4) circle (.05) node[above] {\small $-q_4$};
	\draw [thick, red,decorate,decoration={brace,amplitude=5pt},xshift=0.4pt,yshift=.6cm](-3.3,6) -- (-1.7,6) node[black,midway,yshift=0.5cm] {\small $Q$};
	\draw [thick, red,decorate,decoration={brace,amplitude=5pt},xshift=0.4pt,yshift=.6cm](1.7,6) -- (3.3,6) node[black,midway,yshift=0.5cm] {\small $-\bar{Q}$};
\end{tikzpicture}
\caption{Energetically favored solution for unpaired charges when $Q>\bar{Q}$.}

\end{center}
\end{figure}

The free energy of this solution as a function of the parameter $q_4$ is given by the plot in fig.~\ref{fig:D5q2=q3} for $Q=0.15$ and $\bar{Q}=0.1$.
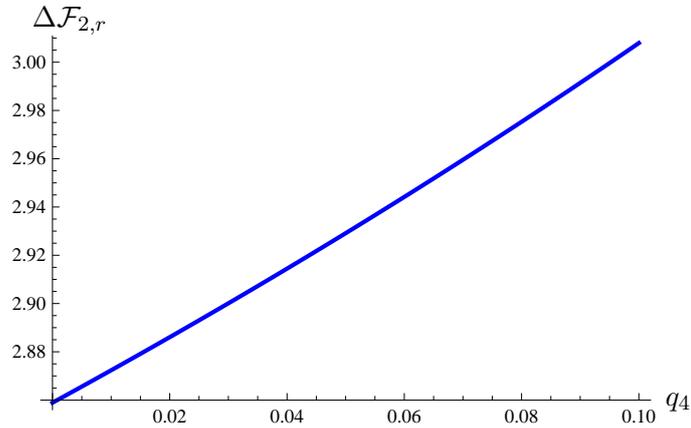
\begin{figure}[h!]
\vskip 2mm
\centerline{
\includegraphics[scale=1]{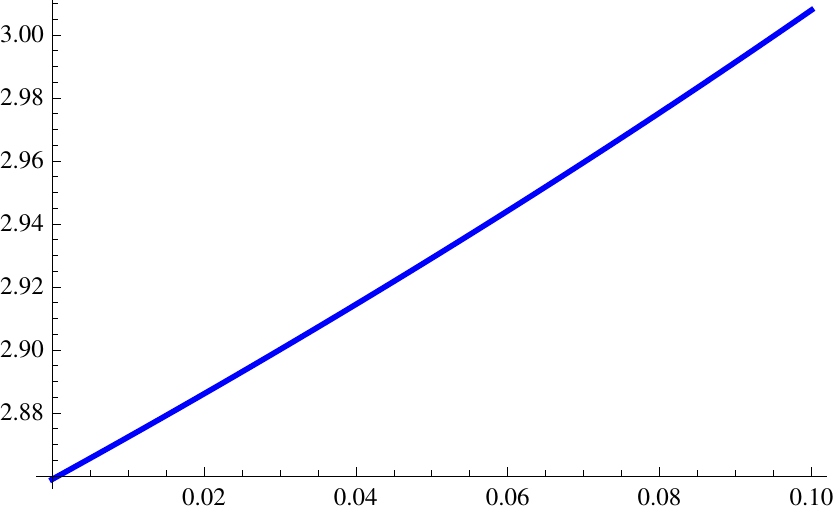}
\begin{picture}(0,0)(0,0)
	\put(-237,151){$\Delta \CF_{2,r}$}
	\put(0,8){$q_4$}
\end{picture}}
\caption{\small Free energy of the solutions when one brane is disconnected and two branes are connected}
\label{fig:D5q2=q3}
\end{figure}

From fig.~\ref{fig:D5q2=q3} it is clear that the lowest free energy is achieved when $q_4=0$ which corresponds to the fact that one anti-D5 brane is represented by a Minkowski embedding.

\item\label{D5-2}
Then we can consider the configuration in which all the branes are disconnected. In this case 
\[
q_1=Q-q_2 ~,~~~~~ q_3=\bar{Q}-q_4
\]
In fig.~\ref{fig:D5q1=q2} we give the free energy of the D5 branes, and of the the anti-D5 brane. 
It is clear from fig.~\ref{fig:D5q1=q2} that the lowest free energy configuration is when both branes
on the same layer have the same charge. The free energy of the complete configuration will be then the sum of the free energy of the D5 and of the anti-D5 layers each with charge evenly distributed over the branes. For the case considered we obtain a free energy $\Delta \CF_{2,r}\simeq 3.26$, which however is higher then the free energy of the configuration $(\ref{D5-1})$.

\begin{figure}[h!]
\vskip 2mm
\centerline{
\includegraphics[scale=.85]{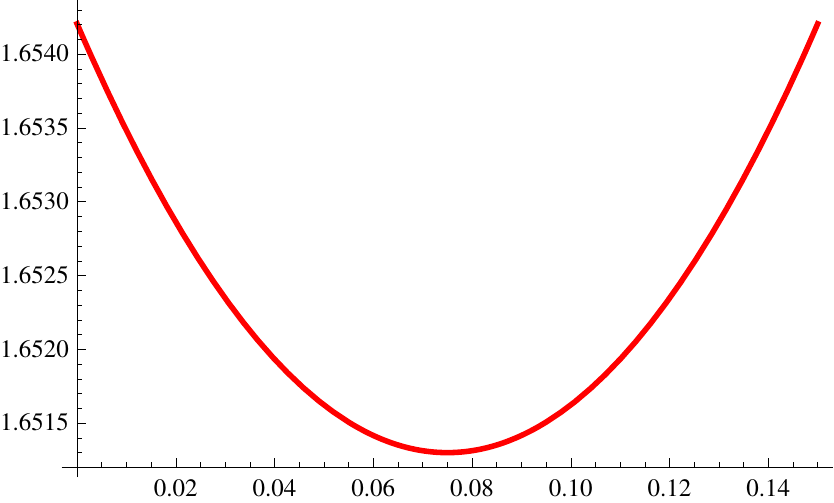}
\begin{picture}(0,0)(0,0)
	\put(-200,126){$\Delta \CF_{2,r}$}
	\put(0,7){$q_4$}
	\put(-105,80){D5}
\end{picture} \hskip 2mm
\includegraphics[scale=.85]{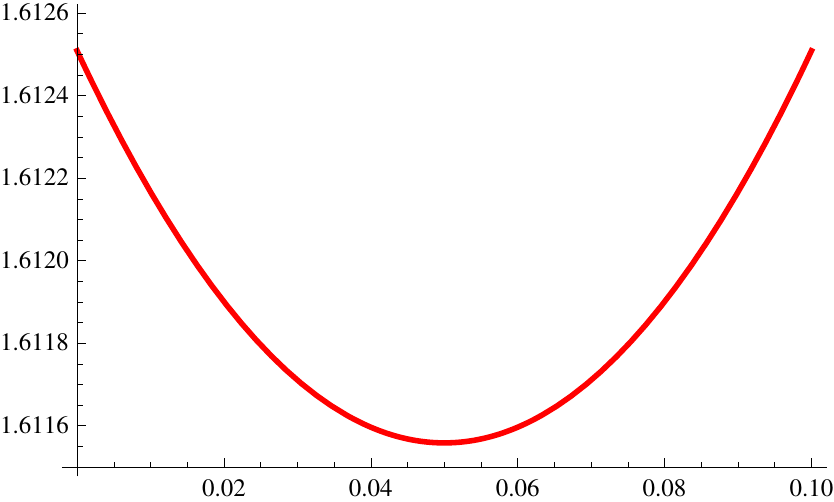}
\begin{picture}(0,0)(0,0)
	\put(-200,125){$\Delta \CF_{2,r}$}
	\put(0,7){$q_4$}
	\put(-105,80){$\overline{\text{D5}}$}
\end{picture}}
\caption{\small Free energy of the solutions when all the branes are disconnected: The energetically favored solution is when they have the same charge.}
\label{fig:D5q1=q2}
\end{figure}
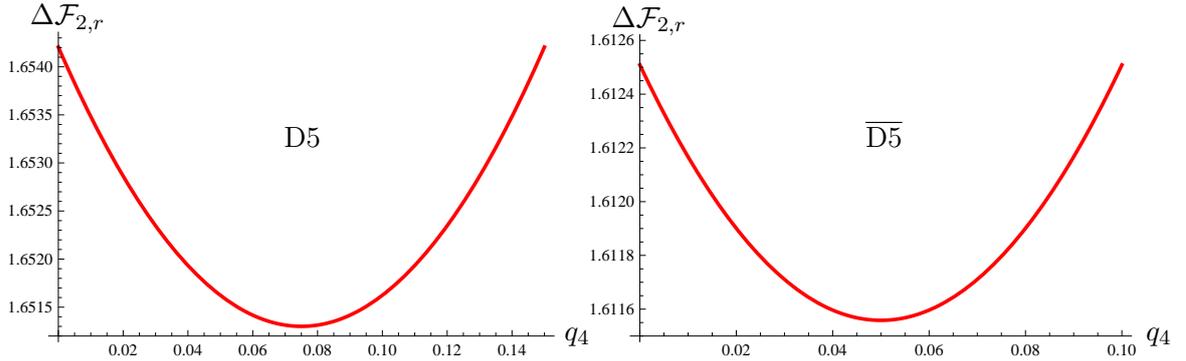

\end{enumerate}

In the special case in which $Q$ and $\bar{Q}$ are equal, $e.g.$ $Q=\bar{Q}=0.15$, there are four possible configurations: Either the branes are all disconnected, or the two pairs branes are both connected, or a brane and an anti-brane are connected and the other are black hole embeddings, or, finally, a brane and an anti-brane are connected and have all the charges $Q$ and $\bar{Q}$, so the rest are Minkowski embeddings. 

When they are all disconnected the physical situation is described in item \ref{D5-2} and the energetically favored solution is that with the same charge. When they are all connected the configuration has the following charges.
\[
q_1=Q-q_2~,~~~~~~ q_3=q_1 ~,~~~~~q_2=q_4~,~~~~~q_3=\bar{Q}-q_4=Q-q_2
\] 
\begin{figure}[h!]
\vskip 2mm
	\centerline{
	\includegraphics[scale=1]{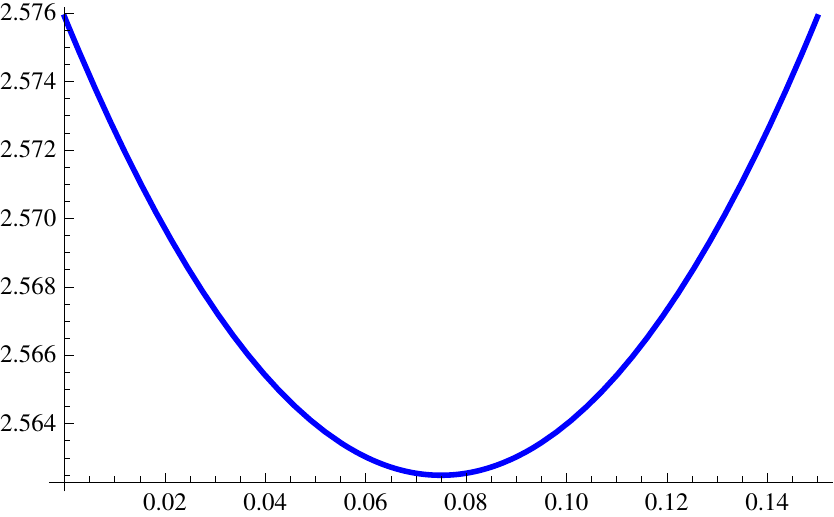}
	\begin{picture}(0,0)(0,0)
		\put(-237,151){$\Delta \CF_{2,r}$}
		\put(0,8){$q_2$}
	\end{picture}}
	\caption{\small Free energy of the solutions when all the branes are connected, the energetically favored solution is when the charge is distributed evenly between the branes and anti-branes.}
	\label{fig:D5q1=q3}
\end{figure}
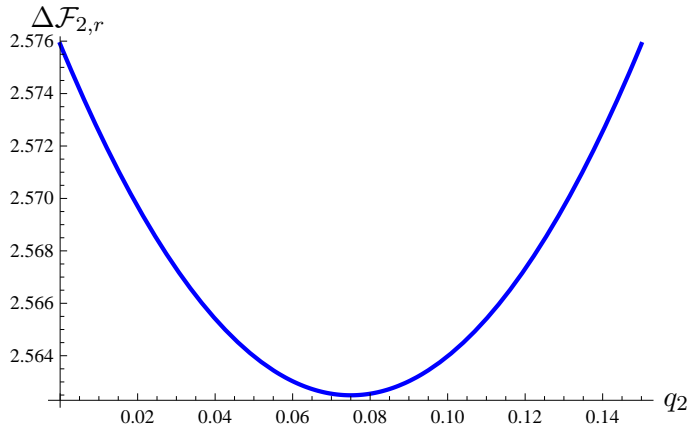
Fig.~\ref{fig:D5q1=q3} shows the free energy of the system of branes and anti-branes when they are all connected, clearly
the energetically most favored solution is that with the charge distributed evenly. This solution has  a lower free energy with respect to the one of all disconnected branes and charges distributed evenly and also with respect to a solution with one brane anti-brane connected system and two black hole embeddings. Since the connected solutions for $L\lesssim 1.357$ and $q=0$ are always favored with respect to the unconnected ones, also the solution with two Minkowski embeddings and one connected solution has higher free energy with respect to the one with two connected pairs.

Summarizing for $Q=\bar{Q}$ the energetically favored solution is the one with two connected pairs and all the charges are evenly distributed $q_1=q_2=q_3=q_4=Q/2$.

\section{Discussion}

We have summarized the results of our investigations in section 1.  Here, we note that there are many problems that are left for further work.  For example, 
in analogy with the computations in reference \cite{bilayercondensate1.6} which used a non-relativistic Coulomb potential, 
it would be interesting to study the double monolayer
quantum field theory model that we have examined here, but at weak coupling,  
in perturbation theory.   At weak coupling, and in the absence of magnetic field or charge density, an individual monolayer is a defect conformal field theory.
The double monolayer which has nested fermi surfaces should have an instability to pairing.  It would be interesting to understand this instability better.  
What we expect to find is an inter-layer condensate which 
forms in the perfect system at weak coupling and gives the spectrum a charge gap.   The condensate would break conformal symmetry and 
it would be interesting to understand how it behaves under renormalization.  

The spontaneous nesting deserves further study.  It would be interesting to find a phase diagram for it to, for example, understand
how large a charge miss-match can be.    

Everything that we have done is at zero temperature.   Of course, the temperature dependence of various quantities could be 
of interest and it would be interesting (and straightforward)  to study this aspect of the model.   

It would be interesting to check whether the qualitative features which we have described 
could be used to find a bottom-up holographic model of double monolayers, 
perhaps on the lines of the  one constructed by Sonner \cite{Sonner:2013aua}.  

In references \cite{Kristjansen:2012ny} and \cite{Kristjansen:2013hma} it was shown that, when the filing fraction $\sim q/b$  of a D5 brane gets
large enough, there is a phase transition where it is replaced by a D7 brane,  We have not taken this possibility into account in the present paper
and we should therefore avoid this regime.  This means that, for a charge density $\sim q$ and a magnetic field $\sim b$ on the D5 or anti-D5
brane world-volumes, we must restrict ourselves to the regime where $q/b$ is small, so that the filling fraction, which was defined as
$\nu=2\pi\rho/NB$ remains less than approximately 0.4.  It would be very interesting to study whether the mechanism of references  \cite{Kristjansen:2012ny} and \cite{Kristjansen:2013hma} also occurs in the double monolayer.

\acknowledgments The work of G.W.S and N.K, is supported in part by the Natural Sciences and Engineering Research Council of Canada.   G.W.S. acknowledges
the kind hospitality of the University of Perugia, where part of this work was done.


\addcontentsline{toc}{section}{References}
\providecommand{\href}[2]{#2}\begingroup\raggedright
\endgroup
\end{document}